\RequirePackage{fix-cm}
\documentclass[smallextended]{svjour3}       
\smartqed  
\usepackage{graphicx}
\usepackage{color}
\usepackage{hyperref}
\usepackage{tabularx}
\usepackage{amsmath}
\usepackage{multirow}
\usepackage{multicol}
\usepackage{tabularx}

\DeclareMathOperator{\logit}{logit}

\definecolor{aqua}{rgb}{0.0, 1.0, 1.0} 
\definecolor{aquamarine}{rgb}{0.5, 1.0, 0.83} 
\definecolor{cyan}{rgb}{0.0, 1.0, 1.0} 
\definecolor{darkcyan}{rgb}{0.0, 0.55, 0.55} 
\definecolor{darkorange}{rgb}{1.0, 0.55, 0.0} 
\definecolor{darkorchid}{rgb}{0.6, 0.2, 0.8} 
\definecolor{darkturquoise}{rgb}{0.0, 0.81, 0.82} 
\definecolor{deepskyblue}{rgb}{0.0, 0.75, 1.0} 
\definecolor{jade}{rgb}{0.0, 0.66, 0.42} 
\definecolor{pumpkin}{rgb}{1.0, 0.46, 0.09} 

\begin{document}

\title{Take a deep breath. 
Benefits of neuroplasticity practices for software developers and computer workers 
in a family of experiments 
}

\titlerunning{Take a deep breath.}        

\author{Birgit Penzenstadler         \and
        Richard Torkar \and
        Cristina Martinez Montes 
}


\institute{B. Penzenstadler \at
              Chalmers$|$Gothenburg University \\
              Lappeenranta Lahti University\\
              Tel.: +46-31-7723126\\
              \email{birgitp@chalmers.se}        \\
              orcid 0000-0002-5771-0455
           \and
           Richard Torkar \at
              Chalmers$|$Gothenburg University \\
              \email{torkar@gu.se}\\
              orcid 0000-0002-0118-8143
        \and
           Cristina Martinez Montes \at
              Chalmers$|$Gothenburg University \\
              \email{montesc@chalmers.se} \\
              orcid 0000-0003-1150-6931
}

\date{Received: date / Accepted: March 2nd 2022}

\maketitle


\begin{abstract}
Context. Computer workers in general, and software developers specifically, are under a high amount of stress due to continuous deadlines and, often, over-commitment. 

Objective. This study investigates the effects of a neuroplasticity practice, a specific breathing practice, on the attention awareness,  well-being, perceived productivity, and self-efficacy of computer workers.

Method. We created a questionnaire mainly from existing, validated scales as entry and exit survey for data points for comparison before and after the intervention. The intervention was a 12-week program with a weekly live session that included a talk on a well-being topic and a facilitated group breathing session. During the intervention period, we solicited one daily journal note and one weekly well-being rating. We replicated the intervention in a similarly structured 8-week program.
The data was analyzed using Bayesian multi-level models for the quantitative part and thematic analysis for the qualitative part.

Results. The intervention showed improvements in participants' experienced inner states despite an ongoing pandemic and intense outer circumstances for most. Over the course of the study, we found an improvement in the participants' ratings of how often they found themselves in good spirits as well as in a calm and relaxed state. 
We also aggregate a large number of deep inner reflections and growth processes that may not have surfaced for the participants without deliberate engagement in such a program.

Conclusion. The data indicates usefulness and effectiveness of an intervention for computer workers in terms of increasing well-being and resilience. Everyone needs a way to deliberately relax, unplug, and recover. Breathing practice is a simple way to do so, and the results call for establishing a larger body of work to make this common practice.

\keywords{Sustainability \and Resilience \and Neuroplasticity \and Breathing \and Longitudinal Study}
\end{abstract}

\section{Introduction}
\label{intro}
Physical, mental, and emotional resilience is necessary for taking good decisions under pressure, staying healthy, and experiencing a good quality of life~\cite{herrman2011resilience}. Resilience is specifically relevant for software engineers in comparison to other knowledge workers as 1) they develop some of the most complex systems in the world where they have to combine computational thinking (e.g., intellectually taxing divide and conquer)~\cite{smith2021computational} with systems thinking (e.g., context dynamics and side effects)~\cite{easterbrook2014computational}, 2) they tend to work either in isolation or in intense team environments~\cite{cockburn2006agile}, often globally distributed~\cite{herbsleb2001global}, and 3) they need empathy and relational skills to communicate effectively with colleagues and clients\footnote{as shown in the  medical field~\cite{derksen2013effectiveness}}, and to put themselves into the shoes of future users and build shared meaning~\cite{broome1991building}. Empathy for connecting with others~\cite{riess2017science} requires self-awareness (which has been linked to productivity~\cite{meyer2019fostering}), and may be at risk due to increased connection via technology~\cite{konrath2013empathy}. 

Yet, software developers (and other computer workers) tend to live high-paced work lives and this comes with long-term consequences for their health~\cite{ostberg2020methodology} and happiness~\cite{graziotin2018happens}. Sleep deprivation is often worn as a badge of honor~\cite{fucci2018need}. In addition, the pandemic has taken a toll on well-being and productivity~\cite{ralph2020pandemic}. For example, people tend to experience lower motivation, productivity and commitment while working from home in a disaster situation~\cite{donnelly2015disrupted}. 

Our survey on `Healthy Habits in Software Engineering' attached to an IEEE Blog article~\cite{penzenstadler2020blog} showed that the number one method named by respondents to counteract perceived stress was physical activity, whether as workout or team-sport or recreational activity. 
While movement is a valuable way to decrease perceived stress and to relax the body, there are two limitations to it: first, not everybody may be able to incorporate physical exercise into their routine due to bodily limitations, and second, restrictions due to  recent developments prevented most team sports for extended periods of time.
Consequently, alternative or additional practices to take care of mental and emotional well-being promise to significantly enhance the overall perception of well-being.

In this article, we explore the use of neuroplasticity practices, more specifically, the use of a breathing practice, in terms of its benefits for software developers and computer workers. 
When looking at the general list of practices to enhance well-being and resilience and to relax and release stress from the body, these can be categorized into (i) movement practices (e.g. Yoga Asana, Tai Chi, Qi Gong), (ii) mental practices (e.g. Meditation, Contemplation), and (iii) breathing practices (e.g. Wim Hof, Pranayama, Rebirthing Breathwork, Holotropic Breathwork). 
The listed movement practices are fairly accessible in terms of physical capabilities compared to other more athletic forms of movement. They may still require four functioning limbs, the confidence to practice them with a group, or an environment that ensures sufficient feedback mechanisms to make sure the exercises are carried out correctly. 
The referenced mental practices offer simple beginner methods and techniques --- however, simple does not equate to easy in this case. Consequently, people who already struggle with stress and/or anxiety and worries may be overwhelmed with the request for determination to sit down quietly and calm their mind.
Therefore, we chose a breathing practice as mode of intervention for our investigation. Specifically, one that was not physically or mentally challenging for the aforementioned reasons.

To this end, we designed an intervention with a weekly live group practice that was held online and followed up by the reflective practice of journaling. In addition, we used surveys for collecting quantitative data.

\textbf{Research Questions:} We answer three main research questions on what changes are observable over the course of the breathwork program in the participants' (1) mindfulness attention awareness and daily perceptions of life, (2) well-being, and (3) productivity and self-efficacy.
\\
\textbf{Contribution:} We provide the first empirical study of the ``Pranayama Vyana Vayu'' breathing practice, which is also the first empirical study on breathwork within engineering. The results indicate usefulness and effectiveness as intervention for computer workers to increase well-being and resilience.

In Section~\ref{sec:bg}, we explain the background of the study and related work. Section~\ref{sec:interv} provides details about the intervention, research questions, applied method and research design. Section~\ref{sec:a} describes the analysis. Section~\ref{sec:r} presents the results. Section~\ref{sec:d} explores these in discussion and Section~\ref{sec:c} concludes with a summary and an outlook. Section~\ref{sec:da} points to the open data archive where the scripts and data have been made available. The appendices include the survey instruments (App.~\ref{app:instr}), model designs (App.~\ref{app:modeldesigns}), and detailed findings (App.~\ref{app:detailedfindings}).

\section{Background}
\label{sec:bg}
In this section, we present the context of stress in software development, the background on breathing practices, the theory of mindfulness and the related work on well-being and resilience in software development and IT work. \\
As we introduce a number of concepts relevant to this study, we provide an overview of the most important terms summarized in a table at the end of this section.

\subsection{Context: Stress in Software Development and IT Work}

Stress factors have a negative influence on cognitive task performance~\cite{medvedyk2019influence}, and lead to burnout~\cite{maudgalya2006workplace,marek2017professional}. Contributing to that, over-scheduling and double-booking have been signs of progress and belonging for two decades. Progress equals fast, and fast equals success, which is a recipe for addiction~\cite{brown2014speed}. 
In addition to the known effects of stress on software quality as evaluated by Akula and Cusick~\cite{akula2008impact}, Amin et al.~\cite{amin2011software} found that occupational stress also negatively affects knowledge sharing, which leads to long-term detrimental effects for software systems development, particularly in global development settings.

Fucci et al. looked into the effects of all-nighters for software developers, as they are often willing to work late for project deadlines because ``forgoing sleep appears to be a badge of honor in the programmers and start-up communities''~\cite{fucci2018need}. Sleep deprivation and disrupted circadian rhythms may lead to adverse metabolic consequences~\cite{buxton2012adverse}, all the way up to increasing the risk for developing cancer~\cite{haus2013shift}. 
The effects of sleep deprivation are clearly negative, and stressed software engineers report a decreased quality of sleep~\cite{akula2008impact}, which in turn negatively impacts health~\cite{hafner2017sleep}. 
This also leads to economic losses, recognized in the US, but also United Kingdom, Japan, Germany, and Canada~\cite{hafner2017sleep}, estimated to between \$280 billion and \$411 billion for the US in 2020, depending on the scenario, and between \$88 billion and \$138 billion for Japan.
Consequently, the potential benefits of the practice evaluated in the study at hand could help improve quality of sleep and decrease the respective physical and mental health consequences.
Lavallée and Robillard~\cite{lavallee2015good} found in a ten month study that many decisions made under the pressure of certain organizational factors negatively affected software quality, which further motivates our goal to increase stress resilience for people in this line of work.

Ostberg et al.~\cite{ostberg2020methodology} show a methodology of how to physiologically evaluate the stress that software developers are under, so that interventions against the stress and its long-term consequences for health can be empirically measured. We are using their self-efficacy instrument in the study at hand.

\subsection{Background: Breathing practices}

\paragraph{Origins: Yogic Pranayama.}\label{sec:pranayama}
The origins of traditional breathing practices, also known as pranayama, are found in the Vedic scriptures that date back to ~\cite{panikkar1994vedic}. The ``Vedas'' are regarded as the world's oldest piece of literature. 
They are the basis of Ayurveda (Science of life) and Yoga (Union of body, mind and spirit).
Yoga made its way into the West during the 20th century, with a huge increase in popularity first in the sixties and seventies and then over the last two decades. One of the ways yoga is practiced is pranayama, composed by the two words `prana' (life force) and `ayama' (that which animates). So what gets somewhat lazily translated as breathing practices are energy practices that serve to increase and adapt the energy flow in a practicing individual. 

\paragraph{Breathwork.}
The term breathwork has been used as a synonym in the West by a wide variety of teachers, so we briefly introduce it at this point.
Specifically the adaptation of energy enhancement and balancing via means of breathing practices has been popularized outside of its yogic origins in the West since the 1960s, by researchers and teachers like Stanislav Grof and Leonard Orr. Stanislav Grof discovered that a specific breathing pattern that he named Holotropic Breathwork produced similar effects as the consumption of LSD (after he had accidentally discovered LSD while testing drugs in the lab for a pharma company, and the substance was proclaimed illegal later on). Leonard Orr coined the Rebirthing Breath after going through an awakening experience in a sensory deprivation floating tank. Both patterns work with circular breathing, which means there are no breaks in between in-breath and out-breath, which can lead to a strong energetic stimulation of the body that can trigger cognitive and emotional experiences. There are many other forms and other accomplished and experienced facilitators like David Elliot~\cite{elliot} and Dan Brule~\cite{brule2017just} who have studied in depth, guided thousands of participants, and pass the knowledge on globally.
In the article at hand, we build on the lineage passed down by the Breath Center~\footnote{\url{http://www.thebreathcenter.com}} where the first author become a certified practitioner in 2019.

\paragraph{The Neuroscience and Empirical Benefits of Breathing Practices.}
The vagus nerve is the largest cranial nerve in our body and a vital player for the parasympathetic nervous system~\cite{squire2012fundamental}. Our nervous system is in a fight-or-flight response during stress (of any kind, be it physical, mental, or emotional) and to recover more quickly from stress it is vital to tone the vagus nerve~\cite{nestor2020breath}. When we experience stress, the kidneys release adrenaline, which gets transported to the brain via the vagus nerve, where it gets compared and stored to memory.
The breathing practice used in the intervention of the article at hand gently resets the nervous system and thereby provides the grounds for responding to life from a resourced place as opposed to fight-or-flight.

Several studies have been carried out related to the breathing practice of Sudarshan Kriya trained by The Art of Living\footnote{\url{https://www.artofliving.org}}, e.g., Seppala et al.~\cite{seppala2020promoting} address the decline in mental health on U.S. university campuses by examining the effects of three interventions: Sudarshan Kriya breathing (``SKY''; N = 29), Foundations of Emotional Intelligence 
(``EI''; N = 21) or Mindfulness-Based Stress Reduction (``MBSR''; N = 34), with SKY showing the greatest impact, benefiting six outcomes: depression, stress, mental health, mindfulness, positive affect and social connectedness. 
Sharma et al.~\cite{sharma2015rhythmic} explored the topic of the same SKY breathing practice and found positive immunological, biochemical, and physiological effects on health (N = 42). Walker and Pacik~\cite{walker2017controlled} showed a reduction of Post-Traumatic Stress Disorder in Military Veterans (three cases). Brown et al.~\cite{brown2005sudarshan} report it to be used successfully in the treatment of stress, anxiety, and depression.

The family of experiments at hand provides a first empirical evaluation in a related technique and thereby serves as comparative data point as well as confirmatory research for the benefits of breathing practices in general.

\subsection{Theory: Mindfulness and Mindfulness Attention Awareness}\label{sec:mindfulness}

William James (1911/1924), who studied consciousness, was not sanguine about the usual state of consciousness of the average person, stating, ``Compared to what we ought to be, we are only half awake''~\cite[p.~237]{james1911memories}) Based on this and according to~\cite{brown2003benefits}, mindfulness is inherently a state of consciousness. A direct route through which mindfulness may enhance well-being is its association with higher quality or optimal moment-to-moment experiences~\cite{brown2003benefits}. 

Mindfulness-based practices have been studied and evaluated in research since 1979 by Kabat-Zinn~\cite{kabat2003mindfulness}, who developed a clinical 8-week Mindfulness-based Stress Reduction (MBSR) program that has been successfully replicated all over the world, including in correctional facilities~\cite{samuelson2007mindfulness}. 
Kabat-Zinn clarifies that meditation is a direct and very convenient way to cultivate greater intimacy with your own life unfolding and with your innate capacity to be aware~\cite{kabat2021meditation}. 
The objective was to offer an environment with methods for facing, exploring, and relieving suffering at the levels of both body and mind, and understanding the potential power inherent in the mind-body connection itself in doing so~\cite{kabat2003mindfulness}. Clinically proven results include positive affect with regard to emotionally stressful situations as well as increased immune system response~\cite{kabat2003mindfulness}. 

Westen~\cite{westen1996psychology} defined that consciousness encompasses both awareness and attention. \textit{Awareness} is considered the background ``radar'' of consciousness, continually monitoring the inner and outer environment, and one can be aware of stimuli without them being at the center of attention. \textit{Attention} is defined as a process of focusing conscious awareness, providing heightened sensitivity to a limited range of experience.

Thus, \textit{mindfulness attention awareness}, defined as a concept by Brown et al. as ``present-centered attention–awareness''~\cite[p.~824]{brown2003benefits} - which can be explained as being conscious of being mindful - plays a broad and important role in self-regulation and emotional experience, which also impacts work and productivity. Therefore, we included the standard (validated) instrument of the Mindfulness Attention Awareness Scale~\cite{brown2003benefits} into the study at hand.
The MAAS was developed to examine empirical links between mindfulness and well-being, and is focused on the presence or absence of attention to and awareness of what is occurring in the present rather than on attributes such as acceptance, trust, empathy, or gratitude~\cite{brown2003benefits}.

By evaluating the change in mindfulness attention awareness is one of the variables looked at, we investigate whether the used breathing practices contribute to a potential improvement.

\subsection{Related work: Well-being and resilience for engineers, software developers and IT workers}

Bernardez et al.~\cite{bernardez2014controlled,bernardez2018experimental} performed experiments showing that the practice of mindfulness significantly improves conceptual modeling efficiency and improves effectiveness. The authors pointed out that specifically introverts may benefit, and the software field is dominated by introverts~\cite{capretz2003personality}. 

Graziotin et al.~\cite{graziotin2018happens} investigate the happiness of developers and found consequences of unhappiness that are detrimental for developers’ mental well-being, the software development process, and the produced artifacts. They use the SPANE instrument~\cite{diener2009new} to measure differences in the perception of positive and negative affect in experiential episodes, which is also used in the study at hand.

Rieken et al.~\cite{rieken2019mindfulness} explore the relationship between mindfulness, divergent thinking, and innovation, specifically among engineering students and recent engineering graduates in two studies. In the first, they looked at the impact of a 15-minute mindfulness meditation on divergent thinking performance among 92 engineering students at Stanford University. Previous studies have shown that a single meditation can improve idea generation in general student populations. Engineering students who reported higher baseline mindfulness performed better on the divergent thinking tasks. 
The impact of a single 15-minute mindfulness session on divergent thinking performance was to improve the originality of ideas in the idea generation task, but not to impact the number of ideas students came up with in the idea generation task or the engineering design task.

In the second study, they look at the relationship between mindfulness and innovation in survey results from 1400 engineering students and recent graduates across the U.S. from the longitudinal Engineering Majors Survey~\cite{sheppard2010exploring}, to measure baseline mindfulness and confidence in one’s ability to be innovative. Baseline mindfulness predicted innovation self-efficacy across the engineering sample, where a mindful attitude was the strongest predictor of innovation self-efficacy. 
This suggests that the more essential component is the attitude with which you pay attention – or whether you have an open, curious, and kind attitude, often referred to as “beginner’s mind”. 

The only other work we were able to identify up to now that targets breathing practices in IT is den Heijer et al.~\cite{den2017don} who performed a controlled experiment with agile teams that practiced three minutes of a breathing technique for a month at the beginning of every Daily Scrum meeting. The participants perceived the practice as useful, and statistically significant improvement was reported on some of the dimensions in the groups performing an exercise that included listening, decision-making, meeting effectiveness, interaction, and emotional responses.
In contrast, our study works with individuals instead of teams, and a breathing practice designed to support long-term well-being as opposed to short-term situational interventions.

The aim of the study at hand is to further contribute to the body of knowledge of how to decrease stress and increase well-being and resilience for software developers and IT workers.

\paragraph{}Table~\ref{tab:concepts} summarizes the most important concepts introduced in this section along with their differentiation.



\begin{table}[h]\footnotesize
    \centering 
    \begin{tabular}{p{4.5cm}|p{6.5cm}}
        \textbf{Concept} & \textbf{Definition} \\\hline
        Mindfulness & a state of consciousness, the practice of purposely bringing one's attention in the present moment without evaluation~\cite{kabat2021meditation}\\
        Attention & the behavioral and cognitive process of selectively concentrating on a discrete aspect of information~\cite{james2007principles}\\
        Attention capacity & Amount of cognitive resources available within a person~\cite{kahneman2011thinking}\\
        Awareness & the quality or state of being aware, knowledge and understanding that something is happening or exists~\cite{chalmers1996conscious}\\
        Mindfulness Attention Awareness & present-centered attention–awareness~\cite[p.~824]{brown2003benefits}\\
        Stress & non-specific response of the body to a demand~\cite{selye1956stress}\\
        Well-being & what is non-instrumentally or 
ultimately good for a person~\cite{crisp2001well}\\
        Resilience & positive adaptation, or the ability to maintain or regain mental health, despite experiencing adversity~\cite{herrman2011resilience}\\
        Self efficacy & People's beliefs about their capabilities to produce effects.~\cite{bandura1994self}\\
        \hline
    \end{tabular}
    \caption{Summary of the most important terms relevant for the study at hand}
   \label{tab:concepts}
\end{table}

\section{Research Design}\label{sec:interv}

Why use a breathing practice as a central technique to increase well-being and resilience?
Research shows that \textit{meditation} is great for enhancing emotional resilience and a healthy stress response~\cite{evans2009using,creswell2014brief}. However, people have been restless at home, so for many the idea of meditating can trigger additional anxiety or restlessness, which counters the intention for the exploration of this study.
Research shows if we can engage in \textit{deliberate movement} where the mind gets focused on a task, this has calming effects, for example in yoga or running, but also extreme sports~\cite{kotler}. Explosive and\slash or high-intensity intervals can be great for exerting, then recovery, where the recovery part is crucial for the benefit of the overall activity, physically and mentally. Consistent activity and\slash or moderate intensity is sometimes better for lowering stress because if the body is already under a high level of stress hormones, the additional input can burden the body further, especially if the individual is not used to increased levels of physical activity. It depends on personality and physiology whether we relax better after exertion, or by means of moderate activity~\cite{kotler}. However, most importantly for the study at hand, it requires certain physical abilities by the participants whereas we wanted to make this \textit{intervention accessible for everyone}. And everyone has to breathe.

Our brains and bodies learn through repetition and establishing habits, and ``Your actions today become your brain's predictions for tomorrow, and those predictions automatically drive your future actions''~\cite[p.~82]{feldmann2020brain}. Consequently we chose for the intervention to last long enough to establish a new habit.

The program is unique to the best of our knowledge in adapting breathwork practice and framing topics specifically for a computer worker and software engineering background. The closest program we came across is the mindfulness program by Google~\cite{tan2012search}, which is based on meditation.

\subsection{The intervention: Rise 2 Flow program}
We created a program to help build mental and emotional resilience and increase well-being.\footnote{\url{https://www.twinkleflip.com/rise-2-flow/}} This program is built around a specific yogic breathing practice, a so-called Pranayama (see Sec.~\ref{sec:pranayama}) that affects Vyana Vayu (the `wind of the nerves') or nervous system. It is a three-part breath through the mouth that is practiced laying down. The first part is an inhale in the belly, the second one an inhale into the chest, and the third part is a complete exhale. This specific pattern triggers a release in the parasympathetic nervous system and thereby helps to deeply relax. The practice is very gentle and therefore easy to use for people who are new to this type of modality.
The first author is a certified facilitator of this particular technique.

\begin{figure}
    \centering
    \includegraphics[width=\textwidth]{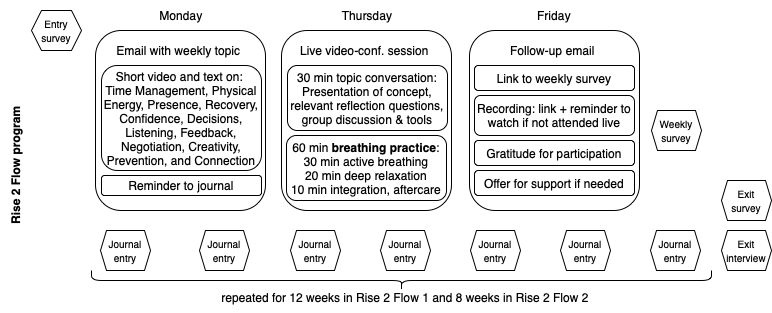}
    \caption{An overview of the Rise 2 Flow intervention}
    \label{fig:rise2flow}
\end{figure}
The main component of the intervention program was the breathing practice. We framed the breathing practice with a weekly topic in the area of self-development. 
The weekly topics were used as framing for two reasons: First, to increase the interest in the study and give the participants an additional buy-in to learn applicable tools (to speak more to the left brain hemisphere oriented thinkers that we tend to have as a majority in software engineering), and to give them these tools such that they could make use of and harness their increased attention awareness and use their energy in more effective and efficient ways.
In the 12-week version (September to December 2020), the topics were Time Management, Physical Energy, Presence, Recovery, Confidence, Decisions, Listening, Feedback, Negotiation, Creativity, Prevention, and Connection. In the 8-week version (January to March 2021), we used the first eight of that list. The choice of topics is based on extensive reading in self development over the past decade and a selection of topics that to us appeared most relevant to the target population in relation to their well-being and resilience at work.
Each of the topics gets primed at the beginning of the week by an email that offers a few questions to reflect upon over a few days until the group breathing practice session. A diagrammatic overview is provided in Fig.~\ref{fig:rise2flow}.
In the practice session, conducted via an online conferencing tool, we start with a conversation around these questions (e.g., ``How much sleep do you get on average?'') to give the participants a bit of time to wind down as they usually come out of an intense workday.
Then we give the instructions for the breathing practice, which is performed for three rounds of seven minutes each with brief relaxation pauses in between, followed by a 20-min relaxation. After the session, we finalize with aftercare suggestions (e.g., to hydrate well) and are available for questions.

There are supplementary materials that the participants can choose to make use of, namely a brief video primer introducing the weekly topic, a guided meditation on the weekly topic, the presentation slides that are used during the live session, and a workbook derived from those slides with a few more reflection prompts to journal on if desired.
For participants who could not make it to the live sessions, we recorded them and made them available for home practice later during the week.\footnote{\url{https://www.twinkleflip.com/rise2flow/} with recording links removed for privacy of the participants.}

We scheduled and administered data collection by means of surveys, in an entry and exit survey (about 20 min), a short weekly survey (5 rating items and a comment), and a daily journal entry (1 rating and a comment). We offered exit interviews, but these are omitted from the result analysis in this article for reasons of space. \\
The invitation to a brief daily journal entry and short weekly surveys were necessary to collect data that would help us understand that comparison points from entry and exit survey and to see a development over time. We do not see them as part of the intervention per se, as such a reflective practice would not necessarily change the wellbeing of participants much by itself (unless specifically prompted, for example, as gratitude practice~\cite{chlebak2013mindfulness}) but rather helps to notice changes that occur (in our case, prompted possibly by the intervention).

We conducted two pilots, which we include here for completeness and also to report a logistic failure in the first one\footnote{Because we report rarely on failure in our discipline, we miss out on collective learning from these failures.}. The first attempt enrolled 30+ participants in Spring but then led to only 10 submissions of the entry survey and 0 submissions of exit survey. The logistic failure was that the first author enrolled students from a course she was teaching and the ethical research guidelines required that she did not know who in the course was participating in the study (to mitigate the risk of coercion) and therefore she could not follow up with participants after sending the initial instructions. In addition, there were no live components, only recordings, so that may have significantly lowered engagement.\\
The second pilot came into existence upon reporting on the failed pilot at a conference (ICT for Sustainability 2020), where researchers in the audience volunteered for a second round. So we set up a second round, this time with direction communication, and an option to participate in weekly live sessions or to listen to recordings. However, the timing was sub-optimal as many participants dropped off over summer (30+ sign-ups, 15 submissions of the entry survey, 3 submissions of the exit survey).

In the two instances from the family of experiments (Rise 2 Flow 1 and Rise 2 Flow 2), direct interaction online was an important component, and the timing was just after the start of the academic year. For Rise 2 Flow 1, 34 completed the study exit survey, and for Rise 2 Flow 2, 33 completed the exit survey (both times out of 100+ sign-ups that we had advertised for globally).

\subsection{Research Questions}
\label{sec:hp}

For reporting on the implementation of the study, we are interested in how participants engage with the program. Specifically, how many sessions they attended or listened to the recording of, 
what their results of the practices were, 
and which practices participants applied in their daily lives.

The research questions for the study at hand are as follows:
\begin{enumerate}
    \item How did participants' mindfulness attention awareness and daily perceptions of their experience of life change over the course of the breathwork program? 
    \begin{enumerate}
        \item Does the intervention bring about change in the participants Mindfulness Attention Awareness? 
        \item How did the daily perceptions of life experience progress over time? 
        \end{enumerate}
    \item What is the observable change in participants' reported well-being over the course of the breathwork program?
    \begin{enumerate}
        \item Is there change in the participants' perceptions of positive and negative experiences? If so, how are their experiences affected? 
        \item Is there change in their psychological well-being? If so, how is it affected?  
        \item Is there change with regard to their positive thinking? If so, how is it affected?  
        \item How does the well-being fluctuate and vary over the course of the course of the intervention? 
    \end{enumerate}
    \item What are the observable changes in participants' perceived productivity and self-efficacy over the course of the breathwork program?
        \begin{enumerate}
        \item Does the intervention lead to change in the participants perceived productivity? If so, how is it affected?  
        \item Does the intervention lead to change in the participants' self-efficacy? If so, how is it affected?  
        \end{enumerate}
\end{enumerate}

The overall expectation is that well-being and resilience increase over the course of the study, evidenced in quantitative and qualitative data as collected in the survey instruments and interviews.\footnote{We decided to not work with hypotheses for the statistical part because  the American Association for Statistics recommends to not use dichotomous hypothesis testing because very seldom when we study complex systems we can derive to such an easy T/F answer~\cite{wasserstein2016asa}. 
In addition, our study uses a combination of qualitative and quantitative data and therefore provides richer answers than T/F.}


\subsection{Population and inclusion criteria}

The target population are people who work in IT and software development. The inclusion criterion is that they spend at least 70 percent of their work time in front of a screen.\footnote{We advertised the study as such and asked participants in the sign-up form, and we have to rely on their self-assessment of that criterion.}
We expected a sample that would include software developers, IT practitioners, IT researchers, IT consultants, faculty, and students. We included the latter as options because we find it highly relevant to address the educational aspect of offering practices early on in career development, not only when all routines have been set in place. 
The participants for both runs of the experiment were recruited across a range of personal and online networks, including the global personal network of the first author, university networks, mailing lists, online spaces, and social media channels. 
We reached out to several hundred colleagues around the globe per email to ask them to promote the study in their courses. We pitched the study live to six courses given by colleagues in Sweden and California. 
We also reached out to our alumni network in several countries per email, consisting of several hundred members. Furthermore, we posted invitations on research community mailing lists (ICT for Sustainability, LIMITS), and advertised in a series of posts on Twitter, LinkedIn, Facebook, Instagram, and several large Slack spaces. There was no compensation for the study, so the only incentive was to learn the breathing technique and practice in a facilitated group.
The video pitch for the study is available here: \url{https://youtu.be/ifdo4-ZCoFM}

While it is not a classical convenience sample because of the number of channels used for broadcasting, it can be seen as an extension thereof~\cite{baltes2020sampling}.

This study was carried out in accordance with the recommendation for experimental guidelines of Chalmers University of Technology with informed consent from all subjects. Because of the informed consent and the non-intrusive nature of the study, no formal ethics committee was required to review the study as per the university's guidelines and national regulations.

\subsection{Instrument design}\label{subsec:instr}

The instruments for the Rise 2 Flow study comprise an entry survey and an exit survey, to be taken before the first practice session and after the last practice session. It encompasses items on attention awareness, positive and negative experiences, psychological well-being, positive thinking, perceived productivity, and self-efficacy, (Sect.~\ref{sec:entry}).

In addition, there is a weekly well-being check-in survey, where participants rate their well-being using five items (Sect.~\ref{sec:weekly}), as well as a daily journal entry (Sect.~\ref{sec:daily}). The instruments are included in App.~\ref{app:instr}.

Answering several calls in the field~\cite{feldt2008towards,lenberg2015behavioral,gren2018standards,wagner2020challenges}, this work adopts validated measurement instruments that come from psychology.

\subsubsection{Entry\slash exit survey}\label{sec:entry}

Our entry survey is composed of several validated instruments in related work, the Mindfulness Attention Awareness Scale (MAAS), the Scale of Positive and Negative Experience (SPANE), the Psychological Well-Being scale (PWB), the Positive Thinking Scale (PTS), a Perceived Productivity instrument (HPQ), and a Self-Efficacy instrument---all of which are introduced and explained in this section.
The exit survey had those same instruments, in order to have a comparison point.

\paragraph{Mindfulness Attention Awareness Scale (MAAS, for RQ1a)} 
Brown and Ryan~\cite{brown2003benefits} presented their scale to validate the benefits of being present by demonstrating the role of mindfulness in psychological well-being under the name `Mindfulness Attention Awareness Scale' (MAAS). The instrument assesses individual differences in the frequency of mindful states over time. Its development began with a pool of 184 items that was subsequently reduced to 55 and then 24 items. 

After exploratory factor analysis, the final version included 15 items. The items are distributed across cognitive, emotional, physical, interpersonal, and general domains. MAAS respondents indicate how frequently they have the experience described in each statement using a 6-point Likert scale from 1 (almost always) to 6 (almost never), where high scores reflect more mindfulness. To control for socially desirable responding, respondents are asked to answer according to what ``really reflects'' their experience rather than what they think their experience should be~\cite{brown2003benefits}. 

It has been widely used in clinical psychology, e.g.,~\cite{shapiro2008cultivating,baer2006using}, behavior assessment, e.g.,~\cite{mackillop2007further}, cognitive therapy, e.g.,~\cite{evans2008mindfulness}, and psychosomatics, e.g.,~\cite{carmody2008mindfulness}.


Most later research has verified its validity, for example Baer et al.~\cite{baer2006using} combined five questionnaire instruments and confirmed MAAS' good psychometric properties. Barajas and Garras confirmed its validity in a large Spanish sample~\cite{barajas2014mindfulness} and Deng et al.~\cite{deng2012psychometric} in China. MacKillop and Anderson~\cite{mackillop2007further} performed a confirmatory factor analysis that supported the unidimensional factor structure of the MAAS in their overall sample.

MAAS has been critized for only one aspect, which is using negative statements in their rating~\cite{hofling2011mindfulness}, which could affect construct validity. Hofling et al.~\cite{hofling2011mindfulness} propose that MAAS can be assessed by both positively and negatively worded items if trait-method models are applied. Their 10-item version MAAS-Short uses five positively and five negatively worded items and is superior to the MAAS with regard to internal consistency, but content validity might be restricted with fewer items.

Consequently, we chose to use the original version of the instrument.

\paragraph{Scale of Positive and Negative Experience (SPANE, for RQ2a)} 
Diener et al.~\cite{diener2009new} proposed a set of related instruments in `New measures of well-being' that includes the Scale of Positive And Negative Experience (SPANE), the scale of Psychological Well-being (PWB), and the Positive Thinking Scale (PTS). In his meta analysis of studies applying Diener et al.'s instruments, Busseri~\cite{busseri2018examining} examines the structure of subjective well-being, and confirms the associations among positive affect, negative affect, and life satisfaction. To measure these different aspects, all three instruments are part of the entry and exit survey in the study at hand. 

The Scale of Positive and Negative Experience (SPANE) elicits a score for positive experience and feelings (using six items), a score for negative experience and feelings (six items), and the two are combined to create an experience balance score. The respondent selects on a Likert scale how often they have experienced the specific feeling over the past month. The scale assesses a broad range of negative and positive experiences and feelings with only twelve items. 

Each item is scored on a scale ranging from $1$ to $5$, where $1$ represents ``very rarely or never'' and $5$ represents ``very often or always.'' The summed positive score can range from $6$ to $30$, and the negative scale has the same range. The two scores are combined by subtracting the negative score from the positive score, and the resulting scores can range from $-24$ to $24$. 
The SPANE is based on the duration during which people experience the feelings, which is beneficial because this aspect of feelings predicts long-term well-being, and it can be better calibrated across respondents. Furthermore, the SPANE is based on feelings that occurred during the previous four weeks, and thus reflects a balance between memory accuracy and experience sampling~\cite{diener2009new}.

The instrument has been mostly supported by later studies, including for example by Jovanovic~\cite{jovanovic2015beyond} wo demonstrated that SPANE is a useful measure of affective well-being. It performs better than the earlier Positive and Negative Affect Schedule (PANAS) by Watson, Clark, and Tellegen~\cite{watson1988development}, in predicting well-being among young adults and adolescents.

\paragraph{Psychological Well-Being (PWB, for RQ2b)} 
Diener et al.'s~\cite{diener2009new} scale of Psychological Well-being (PWB) is a broad measure of a number of aspects of psychological well-being. It assesses meaning, positive social relationships (including helping others and one's community), self-esteem, and competence and mastery.
The PWB provides a good assessment of overall self-reported psychological well-being. While, for the objective of brevity, it does not assess the individual components of psychological well-being described in some theories, it proved to have high internal and temporal reliabilities and high convergence with other similar scales~\cite{diener2009new}.

The Psychological Well-Being scale (PWB) consists of eight items describing important aspects of human functioning ranging from positive relationships, to feelings of competence, to having meaning and purpose in life. Each item is answered on a 1–7 scale that ranges from Strong Disagreement to Strong Agreement. All items are phrased in a positive direction. Scores can range from 8 (Strong Disagreement with all items) to 56 (Strong Agreement with all items). High scores signify that respondents view themselves in very positive terms in diverse areas of functioning.

\paragraph{Positive Thinking (PTS, for RQ2c)}
People's habits of positive thinking are not the sole determinant of happiness as circumstances can influence well-being as well. However, the propensity to positive or negative thinking can influence a person's feelings of well-being, while controlling for environmental circumstances. Thus, Diener et al.~\cite{diener2009new} developed a scale of Positive Thinking (PTS) as a measure of the propensity to view things in positive versus negative terms. 

The Positive Thinking Scale (PTS) is composed of 22 items, where 11 items represent positive thoughts and perceptions and 11 items represent low negative thinking. The 22 items are answered on a yes\slash no format. Negative items are reverse scored with a `no' response counting as a `1'; and for positive items a `yes' response counts as a `1'. After reversing the negative items, the 22 items are added, thus yielding scores that range from 0 to 22~\cite{diener2009new}. 

The authors point out that currently the focus is on attention and interpretation, while taking into account both rumination and savoring would require greater sampling of memories. In addition, a desirable future extension of the scale would be to include thoughts about nonsocial aspects of the world~\cite{diener2009new}.
For the study at hand, our reason for including the scale was that  low PST scores might contribute to explaining variance of well-being scores in otherwise similar contexts. 

\paragraph{Perceived Productivity (HPQ, for RQ3a)}
To assess perceived productivity we used items from the WHO's Health and Work Performance Questionnaire (HPQ) \cite{kessler2003world}, a self-report instrument designed to estimate the workplace costs of health problems in terms of reduced job performance and sickness absence. It was developed because untreated (and under-treated) health problems demand substantial personal costs from the individuals who experience them as well as from their families, employers, and communities~\cite{kessler2003world} and was later validated further as an adequate instrument~\cite{scuffham2014exploring}.

The HPQ\footnote{http://www.hcp.med.harvard.edu/hpq} measures perceived productivity in two ways: First, it uses an eight-item scale (summative, multiple reversed indicators), that assesses overall and relative performance, and second, it uses an eleven-point list of general ratings of participants' own performance as well as typical performance of similar workers.

\paragraph{Self-efficacy (for RQ3b)}
We used the same Self-efficacy instrument used by Ostberg et al.~\cite{ostberg2020methodology} in their work on psycho-biological assessment of stress. The instrument was developed by Jerusalem et al.~\cite{jerusalem1999skala} and based on Bandura et al.'s~\cite{bandura1999self} self-efficacy model.
It is used to assess the individual stress resilience of the participants and encompasses ten items that offer a positively phrased statement on change, challenges or unexpected circumstances which the participant has to rate as ``Not true'', ``Hardly true'', ``Rather true'' or ``Exactly true''.

The study at hand, serves to gain further insights into and potentially confirm the correlation of well-being, positive thinking, and stress resilience.

\paragraph{Personal Data}
We collected information on the participants' country of residence, gender, living situation, occupation, and age. 
The options we offered were: (1) Gender: Man (including cis-man and trans-man), Woman (including cis-woman and trans-woman), Non-binary, and 'prefer not to say'. (2) Living situation: by themselves, with a partner, with their family, in shared housing. (3) Occupation: Student, Faculty, Researcher, Developer, Administrator, IT Services, Manager, Digital Artist, Analyst, Consultant, Retired, Currently not working, and Other.

\subsubsection{Daily journal (for RQ1b)}\label{sec:daily}

The daily journal entry contained three items (plus the participant's alias and the date, for reference):

\begin{itemize}
    \item Which well-being practice did you do today (if any)? Select all that apply: Breathing practice, Yoga postures, Meditation, Nature time, Other.
    \item How was your day? Select a rating from `really bad' (1) to `absolutely great' (10).
    \item Please write about 100 words: What stood out to you today? What caught your attention? What makes you reflect?
\end{itemize}

We intended to have a daily plot over time, and to check the correlation with the practices that were carried out. We were aware that there is a bias in rating as people may not do a well-being practice or even remember to write a journal entry on days where they felt particularly high or low.
The free text gave room for individual reflection and was deliberately prompted in a very open manner.

\subsubsection{Weekly survey: WHO-5 (for RQ2d)}\label{sec:weekly}


The 5-item World Health Organization Well-Being Index (WHO-5) is a short and generic global rating scale measuring subjective well-being. Because the WHO considers positive well-being to be another term for mental health~\cite{jahoda}, the WHO-5 only contains positively phrased items, and its use is recommended by~\cite{bech1999health}. 
The items are: (1) `I have felt cheerful and in good spirits', (2) `I have felt calm and relaxed', (3) `I have felt active and vigorous', (4) `I woke up feeling fresh and rested', and (5) `My daily life has been filled with things that interest me'. 
The respondent is asked to rate how well each of the 5 statements applies to him\slash her\slash them when considering the last 14 days. Each of the 5 items is scored from 5 (all of the time) to 0 (none of the time). 

Topp et al.~\cite{topp20155} performed a recent systematic review on the WHO-5, which included 213 articles from the PubMed and PsycINFO databases. They concluded the WHO-5 has high validity, can be used as an outcome measure balancing the wanted and unwanted effects of treatments, and is a sensitive and specific screening tool. Furthermore, its applicability across study fields is very high.
Consequently, it is a valid choice for the purpose at hand.

We included the weekly survey to be able to observe a development over time. We modified the instrument slightly as we asked to consider the last week, because we administered the survey every week within the 12 weeks of the study. The instrument is included in the replication package~\cite{penzenstadler2021opendata}.

This weekly measurement is relevant for several reasons: 1) To have a longitudinal study that shows the trends over time as opposed to only entry and exit data points, 2) To correlate and validate the insights from the difference in well-being ratings from the entry and exit survey, 3) To investigate the correlation with the development of the pandemic (would mood go down with restrictions increasing?), 4) To check for variance and fluctuations in the well-being as additional indicator for long-term psychological well-being~\cite{houben2015relation}. 

\section{Analysis Procedure}
\label{sec:a}

The previous section covered the research design of the interventions, the study, and the instruments that were administered. This section provides the demographics and the two types of analyses that were performed on the data.
First, a three-way quantitative analysis of the administered instruments. Second, a thematic analysis of answers to the open survey questions. After the analyses, we provide a summary of the results in Sect.~\ref{sec:r}.

\subsection{Demographics}\label{sec:demographics}

For Rise 2 Flow 1, of $137$ sign-ups, $87$ converted to the entry survey, and $34$ completed the study by submitting their exit survey. We collected $1040$ individual journal entries.
For Rise 2 Flow 2, of $169$ sign-ups, $101$ completed the entry survey, and $33$ the exit survey. Participants submitted $616$ individual journal entries.

The high drop out rates have a number of reasons, which are partially known to the first author because of emails kindly sent to her by participants who cared to explain their personal reason for not completing the study. The reasons included the live session being in the middle of get-kids-to-bed time, conflicting commitments for some of the days, and general life stress, which participants did note was a good reason to actually do the sessions but they felt they just could not at the time.

The entry surveys showed an age range from 19 to 58, with a majority of participants in their 20s and 30s.

The occupation was a selection where participants could check all that apply, so the following numbers add up to more than the total number of participants:
Student: 33 (run 1) + 45 (run 2) = 78,
Developer: 6 + 11 = 17,
Researcher: 19 + 42 = 61,
Faculty: 10 + 19 = 29,
Other (Consultant, Analyst, Manager, etc.): 27 + 37 = 64;

Participants joined from all over the world: Argentina,
Austria,
Bangladesh,
Brazil,
Canada,
Costa Rica,
Denmark,
Ecuador,
Finland,
France,
Germany,
India,
Iran,
Ireland,
Italy,
Netherlands,
Mexico,
Poland,
Portugal,
Saudi Arabia,
Spain,
Sweden,
Switzerland,
UK,
US, and
Venezuela.

\subsection{Statistical analysis of instruments}

\begin{figure*}[ht]
    \centering
    \includegraphics[width=\textwidth]{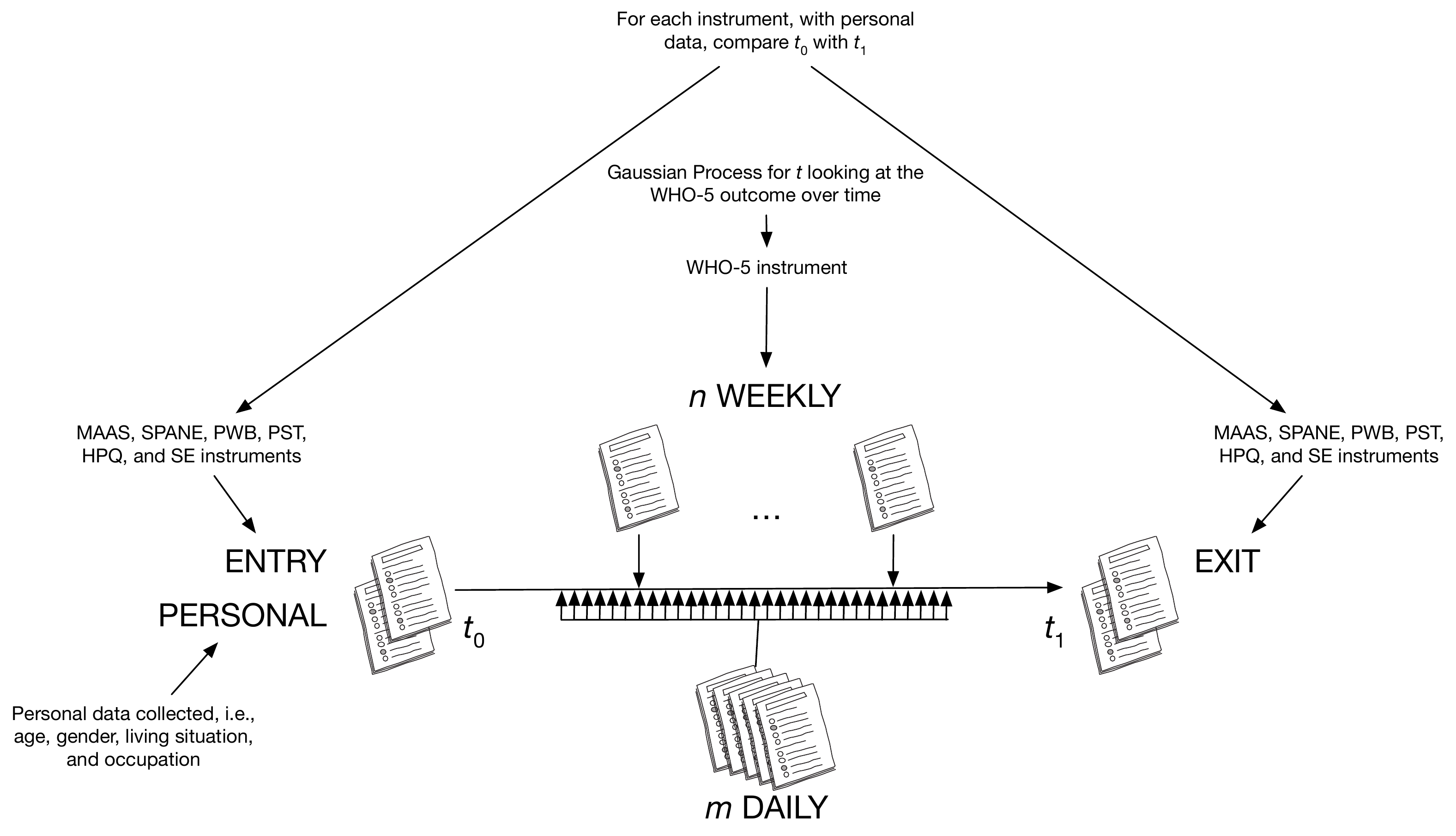}
    \caption{An overview of instruments used in the study. Note that the above was executed twice, once for each run of the experiment.}
    \label{fig:instruments}
\end{figure*}

As presented in Sect.~\ref{subsec:instr}, the study used several instruments for the two experiments. Figure~\ref{fig:instruments} provides an overview of the instruments and at which time they were administered.
A subject was administered an entry survey consisting of six instruments (as presented in the previous section) plus personal data, e.g., current living situation. For each week the subject was administered a weekly survey (the five questions of WHO5) and for each day they answered a daily survey (one question). Finally, upon exiting the study, the subject was administered an exit survey, which was identical to the entry survey.\\
The purpose of the quantitative analysis is to look at how responses change over time (a temporal analysis).\footnote{A replication package can be found at \url{https://github.com/torkar/rise2flow} DOI:10.5281/zenodo.5082388
} This will be done in three ways:

\begin{enumerate}
    \item Temporal analysis for each instrument at $t_0$ vs.\ $t_1$, i.e., entry vs.\ exit.
    \item Temporal analysis of daily trends.
    \item Temporal analysis of weekly trends.
\end{enumerate}
\noindent
In the last case, we will use dummy variable regression estimators (DVRE). The DVRE approach dummy encodes the time variable $t$ and sets an index $0/1$, where $t_0 = 0$ and $t_1 = 1$. In short, each subject (\textsf{ID}) will have two rows where one row are the entry instruments at $t_0$, and one row are the exit instruments at $t_1$. The main reasons to use this approach is: 1. We will see if there is a difference in responses between $t_0$ and $t_1$. 2. If such a difference exists, which predictors, if any, are the main drivers for that difference, i.e., is there a difference in the $\beta$ estimators for each predictor, in each question?\footnote{The same approach was used to encode data for the two experiments that were conducted, i.e., Experiment 1 was coded as $0$ and Experiment 2 was coded as $1$. The main reason for why we chose to encode experiments separately for each subject was to analyze the estimates and their uncertainty; ultimately we needed to ascertain that they did not vary considerably between the two experiments, which could indicate that, e.g., the experiments were not executed in a similar fashion.}

For the first two cases (weekly and daily trends), we will model these using a Gaussian Process ($\mathcal{GP}$). The distribution of a $\mathcal{GP}$ is the joint distribution of all random variables. In short, it is a distribution over functions with a continuous domain, i.e., time in our case. In Bayesian and frequentist multilevel models it is common to model varying intercepts (random effects), which are categorical. However, for continuous values (such as time or space) one needs to use a different strategy. A $\mathcal{GP}$ is such a strategy, i.e., it is a varying intercept approach, but for continuous values.\footnote{Generalized additive models is another approach one could use, however, that approach uses various types of smooth functions instead.}

Before introducing the statistical model design, the next section will present descriptive statistics, dependent and independent variables (and their encoding), and the sample sizes involved in each of our three analyses mentioned above.

\subsubsection{The data and data cleaning}
\begin{figure}[h]
    \centering
    \includegraphics[width=0.9\textwidth]{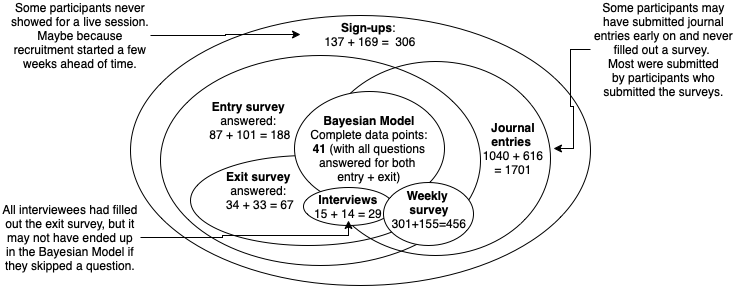}
    \caption{Break-down of collected data and subsets for analysis}
    \label{fig:breakdowndata}
\end{figure}

The entry and exit surveys consisted of six instruments (in parenthesis the number of questions): MAAS ($15$), SPANE ($12$), PWB ($8$), PST ($22$), SE ($10$), and HPQ ($11$). The questions were coded as ordered categorical, i.e., Likert scale, or binary (Yes\slash No answers). 

Concerning the weekly and daily surveys there were five and one question(s), respectively, and the questions were coded as ordered categorical. For the weekly data $98$ subjects answered the survey at least once and some as much as $12$ times, $\tilde{x}=3$ ($N=456$). Concerning the daily data, $111$ subjects answered the survey at least once. The maximum value was $84$, while the median was $\tilde{x}=16$ ($N=1646$).\footnote{The $1646$ differ from the $1701$ in Fig.~\ref{fig:breakdowndata} because the rating of the day was optional and  $1646/1701$ entries were rated.} 

The above are the outcomes.\footnote{Throughout this text we will from now on use the terms outcome and predictor, concerning dependent and independent variables.} The outcomes will be predicted given a number of predictors. An overview of the predictors used in this analysis can be found in Table~\ref{tab:pers_data}.

\begin{table}
    \centering
    \caption{Independent variables (IV) used as predictors.}
    \label{tab:pers_data}
    \begin{tabularx}{\textwidth}{r r X}
    \hline
    IV & Type & Levels \\
    \hline
    \textsf{ID}& factor & Unique for each subject \\
    \textsf{age} & continuous & n\slash a \\
    \textsf{gender} & dichotomous & male $\Vert$ woman \\
    \textsf{occupation} & dichotomous & student $\Vert$ non-students\\
    \textsf{living condition} & categorical & I live by myself $\Vert$ I live in a shared housing $\Vert$ I live with a partner $\Vert$ I live with my family\\
    \hline
    \end{tabularx}
\end{table}

The data cleaning included correcting IDs that the subjects had spelled differently. In some cases we had subjects that only answered one of them; $105$ subjects filled out only the entry survey, while $41$ filled out every single answer for both entry and exit survey.
The Bayesian model required 'complete case analysis'~\cite{mcelreath2018statistical}, i.e. all questions answered, as missing data analysis would have required a causal model for why a participant didn't answer a particular question, which was not feasible in our case due to the number of questions.
While we were curious to investigate where the gender 'non-binary' could be a predictor, the number of answers with that value was too low to result in a valid model. An overview of how the data breaks down for analysis is shown in Fig.~\ref{fig:breakdowndata}.


\subsubsection{Temporal analysis of entry vs.\ exit}
In Appendix~\ref{app:dvrm} a complete specification of the model is listed. Here follows a brief summary of modeling choices; details can be found in the replication package.

A \textsf{Cumulative} likelihood was assumed for the Likert scale questions. In one case (where the outcome consisted of `Yes'\slash`No' answers) the maximum entropy distribution was used (i.e., \textsf{Bernoulli}). Variance between questions in an instrument was modeled using a covariance matrix. Additionally, subject variability (\textsf{ID}) was modeled with adaptive priors to employ partial pooling (more information can be found in Appendix~\ref{app:dvrm} and the replication package).

The question posed for these models was: Given a number of predictors (\textsf{age}, \textsf{gender}, \textsf{occupation}, \textsf{living condition}) is there a difference between $t_0$ (entry) and $t_1$ (exit), when accounting for subject variability (\textsf{ID}).
Prior predictive checks (prior sensitivity analysis) and posterior predictive checks were conducted.\footnote{Additionally, all diagnostics ($\widehat{R}$, ESS, traceplots, E-BFMI, divergences, and treedepth) indicated that the Markov Chains had converged to a stationary posterior distribution.}

To check whether the observed effects correlated with actual participation in sessions instead of only time passing, we also ran models for the predictors of \textsf{total-number-of-sessions-attended}. To observe differences in between participation in live and recorded sessions, we also ran models for the predictors of \textsf{sessions-attended-live} and \textsf{sessions-attended-recorded}.

\subsubsection{Temporal analysis of weekly and daily trends}

In Appendix~\ref{app:GP} a complete specification of the model is listed. Here follows a brief summary of modeling choices; details can be found in the replication package.

Before designing a model, assumptions concerning the data generation process need to be considered. In this study, an information theoretical comparison of possible data generation processes, i.e., \textsf{Cumulative}, \textsf{Continuation ratio}, \textsf{Stopping ratio}, and \textsf{Adjacent-category}, was conducted. The analysis showed that the differences in standard error, between the likelihoods, was fairly large, in comparison to the relative difference in expected log point-wise predictive density. In short, no likelihood showed significantly better out of sample prediction capabilities, when compared to the other likelihoods. Hence, a \textsf{Cumulative} likelihood was assumed for Likert-type questions, while for the questions that were dichotomous the maximum entropy distribution was selected, i.e., \textsf{Bernoulli}.\footnote{Please see the Appendix A in the replication package.}

The statistical model was designed (see Appendix~\ref{app:GP}) with three things in mind. First, the covariance between questions for each subject was modeled employing a covariance matrix. The idea here is that the variability among questions, for each subject, should be captured. Second, the temporal variable (weeks or days) was modeled with a Gaussian Process. Gaussian Process has not been applied in software engineering, as far as we know, but is not uncommon in other disciplines and the concept is perceived as particularly suitable for longitudinal data~\cite{chengV19gp} and variable selection~\cite{pmlr89gp} in a Bayesian context~\cite{vanhatalo13gp}. Finally, when modelling the between-subject variability, partial pooling was used (i.e., self-regularizing priors) to avoid overfitting.

\begin{figure*}
    \centering
    \includegraphics[width=\textwidth]{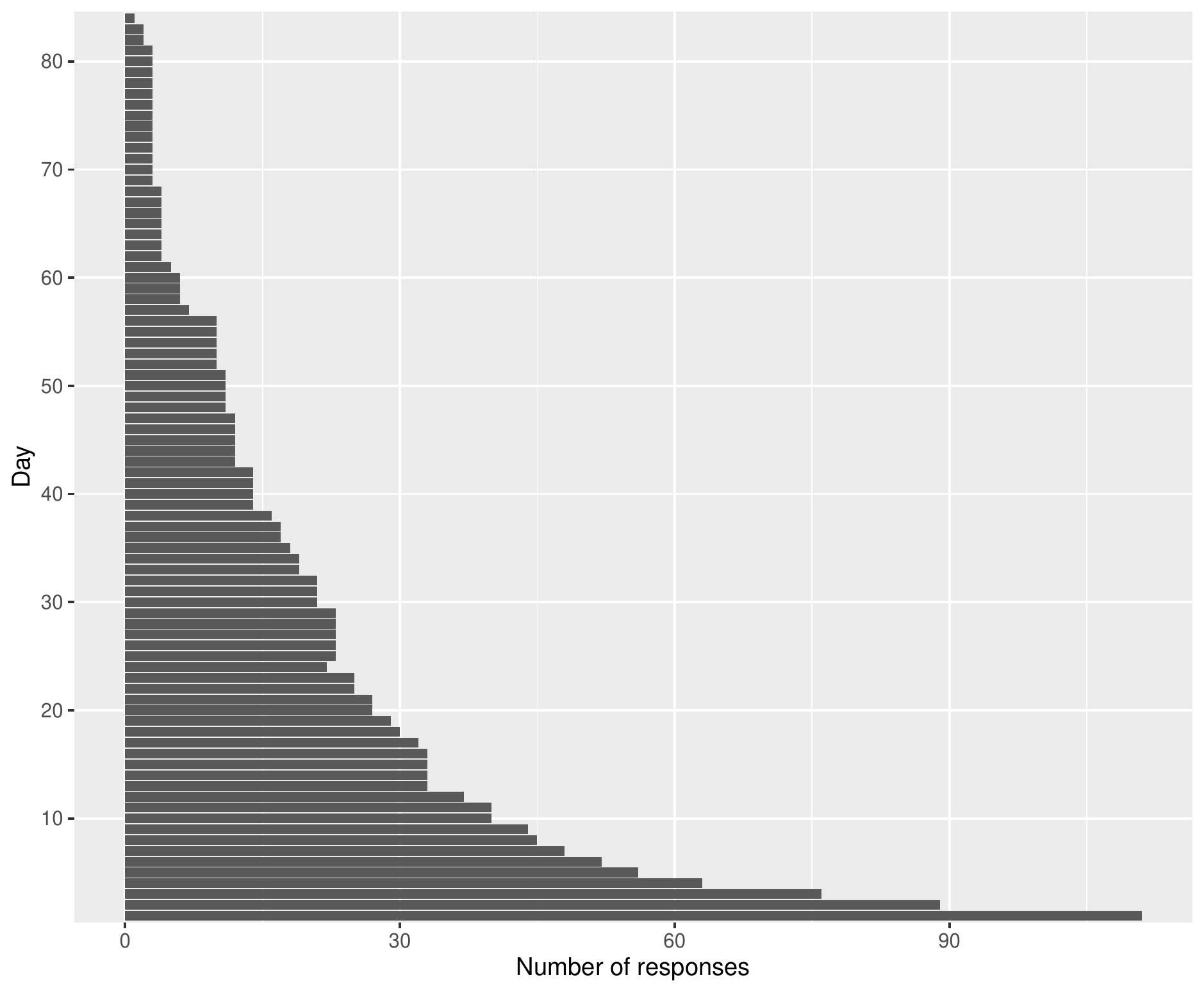}
    \caption{Response rate for each day ($N=1646$). Respondents replied enthusiastically during the first week ($111$ respondents), while on Day $84$ only $1$ respondent replied.}
    \label{fig:ds-daily}
\end{figure*}

Concerning the later design choice, due to us employing a multilevel approach with partial pooling (using a varying intercept for each subject), subjects with a large sample size (answered many times) will inform subjects with a small sample size (answered once or a few times), i.e., the uncertainty will propagate through the model depending on the sample size and we will avoid learning too much from the data (to avoid overfitting). Figure~\ref{fig:ds-daily} shows the challenges researchers face when dealing with response rates in longitudinal studies.\footnote{A survival analysis might be a different method to use in future work.}

In order to handle said threat, and due to the study starting at different time points for each subject, each subject's answer was coded with a time, i.e., $1,\ldots,n$, where $n$ is the last answer they provided. This way a time point in the study, e.g., Week $3$, was the same for all subjects who were still participating in the study at Week $3$. To summarize, the logic was that the intervention, that is, participating in the study, was exchangeable from a statistical point of view, i.e., Week $3$ was the same no matter whether the subject joined the study for the first or the second run. For details on the validity and latency, see App~\ref{app:detailedfindings}, Fig.~\ref{fig:latentscale}.
Prior sensitivity analysis was conducted.\footnote{The priors were uniform on the outcome space (i.e., medians were distributed evenly with large uncertainty). After sampling, posterior predictive checks indicated that each model had learned from the data and washed out the effect of the priors. 
}

As mentioned previously, the model took into account that we had two experiments in this study. By contrasting the underlying latent scale of Experiment 1 and 2 (for each question) we could see that they had overlaying medians and similar shapes.\footnote{See Sects.~2.1.1--2.1.5 in the replication package.} Hence, there were strong indications that the two experiments had been executed in a very similar fashion.

\subsection{Qualitative analysis of instruments: Thematic analysis}

Thematic analysis extracts themes from text~\cite{braun2012thematic}. Coding qualitative research to find common themes and concepts is part of thematic analysis, which is part of qualitative data analysis. 
We present an analysis for the daily journal as well as the entry and exit survey. The coding was performed independently by the first and third author and cross-reviewed.
The weekly survey did not include any open questions.
Table~\ref{tab:codingexample} shows an example of the coding process.

\begin{table}[htbp]
  \centering
  \caption{Example of the coding process}
    \begin{tabular}{|p{15.215em}|c|p{8.57em}|c|}
    \hline
    \textbf{MEANING UNIT} & \textbf{CODE} & \textbf{THEME} & \textbf{SUB-THEME}\\
    \hline
    \textit{``All together, all the awareness of the program has helped me to be more focused and present, enjoying what I am doing, what I don't enjoy still finalize it without falling in the temptation of getting distracted by the first thing that pops in my mind (still happens, but is getting better).''} 
    & Focus
    & Perceived changes in participants 
    & At work \\
    \hline
    \end{tabular}%
  \label{tab:codingexample}%
\end{table}%

For the process of coding, the third author initially coded the data until meaning saturation was reached (``we learned everything from the data that we could)'' to the best of guidance in the state of the art~\cite{braun2021saturate}. The first author audited the coding.
 
\begin{figure}[htb]
    \centering
    \includegraphics[width=\textwidth]{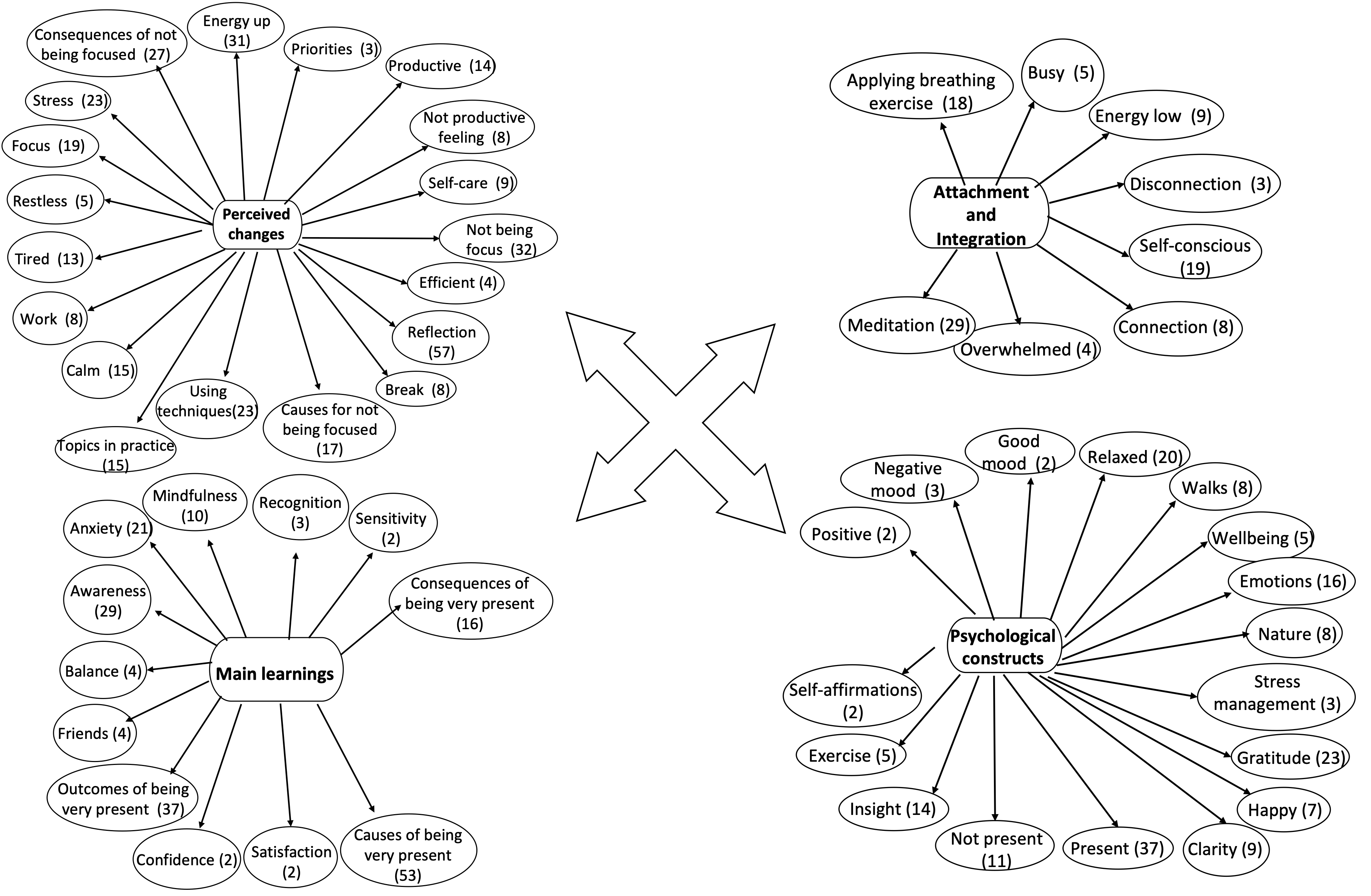}
    \caption{Final Code Map of the Thematic Analysis}
    \label{fig:quantcodemap}
\end{figure}

Figure~\ref{fig:quantcodemap} shows the complete code map of the thematic analysis along with the times that each code was assigned during the coding process.

\section{Results}
\label{sec:r}

\begin{table}[htb]
    \centering
    \begin{tabular}{c|p{6.5cm}|p{2cm}|p{1cm}}
        RQ & Phrasing & Quantitative & Qualitat. \\\hline \hline
        RQ1 & \textbf{How did mindfulness attention awareness and daily perceptions of experience of life change?} & &  \\
        RQ1a & Does the intervention bring about change in the participants Mindfulness Attention Awareness? & negative (MAAS) & positive \\
        RQ1b & How did the daily perceptions of life experience progress over time? & inconclusive \newline (daily) & positive \\\hline
        RQ2 & \textbf{What is the observable change in participants' reported well-being?} &  & \\
        RQ2a & Is there change in the participants' perceptions of positive and negative experiences? If so, how are they affected? & supported \newline (SPANE) & positive \\
        RQ2b & Is there change in their psychological well-being? If so, how is it affected? & supported (PWB) & positive \\
        RQ2c & Is there change with regard to positive thinking? If so, how is it affected? & supported (PTS) & positive \\
        RQ2d & How does the well-being fluctuate and vary over the course of the intervention? & inconclusive (weekly) & positive \\\hline
        RQ3 & \textbf{What are the observable changes in perceived productivity and self-efficacy?} & & \\
        RQ3a & Does the intervention lead to change in the participants perceived productivity? If so, how is it affected? & inconclusive & positive \\
        RQ3b & Does the intervention lead to change in the participants' self-efficacy? If so, how is it affected? & supported & positive \\\hline
    \end{tabular} 
    \caption{Overview of the evidence for answering the research questions}
    \label{tab:results-summary}
\end{table}

In summary (Tab.~\ref{tab:results-summary}), we found a number of quantitative results that were statistically significant, plus a plethora of revelations in the qualitative data that explained some statistical results that had made us wonder, and gave insights into the deep processes of change and growth that some of the participants experienced. 
\\
Where there are quantitative results, we present the estimates that were significant at the arbitrary $95$\% threshold. For each instrument, we will first analyze if there was a difference between questions at $t_0$ vs.\ $t_1$ (i.e., entry vs.\ exit surveys). After that significant effects for the other parameters (for the predictors \textsf{age}, etc.) are presented.
\\
Regarding the qualitative findings to answer the research questions, four themes and four sub-themes were identified presented in Table~\ref{tab:themes} below.

\begin{table}[htbp]
  \centering
  \caption{Themes and sub-themes in response to the RQs}
    \begin{tabular}{|p{7.2cm}|p{3.1cm}|p{0.4cm}|}
    \hline
    \textbf{Themes} & \textbf{Sub-themes} & \textbf{RQ}\\
    \hline
    Main learnings (changes) identified by the participants & & 1\\
    \hline
    Attachment and Integration of techniques in daily life & * Results & 1\\
          & * Actions & 1\\
    \hline
    Psychological constructs modified during the course & & 2\\
    \hline
    Perceived changes in participants & * At work & 3\\
          & * In overall performance & 3\\
    \hline
    \end{tabular}%
  \label{tab:themes}%
\end{table}%

\subsection{Overall Engagement}
Before we answer the individual research questions, we present qualitative findings around the \textbf{overall engagement of participants} with the program to give an insight into the context. 

\paragraph{Attachment and Integration of techniques in daily life---Actions.}
The engagement was variable in each participant. Some people were really committed and put much effort into doing the practices every day, and some other participants could not manage due to their schedules. For example, the quote below shows a strong commitment from this participant based on their need.

\begin{quote}\footnotesize
    \textit{``I will definitely continue with more of these activities (meditation), as they already change my day, mood, and approach towards daily life, routine, and also future plans.''
(participant 14, run 2, journal, Mar 30 2021)}
\end{quote}

On the other hand, there were cases when participants could not find the opportunity during the day to do the practices. However, they found time to write in the journal, as the following quote shows. 

\begin{quote}\footnotesize
    \textit{``Unfortunately, my day was so full, from rising to bedtime, that I didn't have time for any of the practices that remind me to breathe.''
(participant 3, run 1, journal, Nov 26 2020)}
\end{quote}

The commitment of the participants can be seen in the different activities they carried out. Not everyone carried out all the activities at the same time. There were those who attended the live sessions but did not write in the diary. Some participants wrote every day or almost every day and who wrote constantly but not daily. Similarly, there were also a significant number of drop offs.\\
The results of the practices reported by the participants are varied and interesting, as shown in the following quotes.

\begin{quote}\footnotesize
  \textit{``I felt pretty bad so I decided to do the breathing practice. It is interesting to notice that I think I have a new tool to calm after pretty bad days, a tool that does not involve heavy use of alcohol.''
(participant 21, run 1, journal, Sept 28 2020)}  
\end{quote}

This quote is one of the ones that stands out the most, since the participant found in the breathing practice an alternative to calm down instead of the use of alcohol.

\begin{quote}\footnotesize
    \textit{``But I did relieve a growing panic attack with breathing exercises, which felt nice. And of the waking hours, I did feel like I had spent my time more wisely than usual.''
(participant 46, run 2, journal, Jan 30 2021)}
\end{quote}\label{quote:panic}

This participant received a diagnosis of initial depression and burnout; they explain how breathing was the tool that helped them release emotionally.\\
The quote above is also an essential example of the benefits that participants obtained by practicing breathing exercises. This participant wrote that they could control a panic attack and expressed a better use of time.\\
The previous paragraphs describe complex situations in which these participants, through breathing, could find tools that they needed according to their problems.\\
Other results expressed by participants are being more relaxed, in general, and in situations where they previously would not have been; the feeling of being more present at crucial moments of the day; fewer negative thoughts; and ease of letting go. Similarly, some other results were greater focus and better function, more energy, and more centered, a clear mind, calm, peace, gratitude, reflection, or better analysis of situations.

Overall, we observe diversity. 
In terms of \textit{Attendance}, participants came from around the globe, had to deal with time zone differences and work and family situations, and did their best to show up when they could. Some attended only one session live, others almost all of them; some attended only or mainly live, others mainly or only recorded sessions. Calculated from the survey responses, notes from the sessions and views on the online platform, 75\% of the participants who filled out the exit survey had participated in at least 75\% of the sessions. \footnote{Due to the varied attendance, we also ran models with attendance as predictor to make sure the effects we saw over time correlate with attendance and not just time.}
\\
The \textit{Results of their Practice} varied as well. In the exit survey, we asked ``What did you get out of the breathing sessions?'' and the most frequent answers were: 
It is an easy practice to follow (17), I was able to deeply relax during the breathing and in the relaxation period after (22), I remained relaxed after, and felt well rested and recharged the next day (15), I shifted my perception of the world and have interesting insights (9), I feel more present in my body (16), and I have the desire to return to the breathing practice (23).
\\
In terms of \textit{Daily Applied Practices} and records thereof, we received 1032 journal entry submissions. 
Of the daily well-being practices reported in the journal entries, nature time was selected $261$ times ($32.6$\%), followed by meditation ($240$\slash $30$\%), breathing practice ($227$\slash $28.4$\%) and yoga poses ($82$\slash $10.3$\%).
Other practices respondents listed include dancing, offline time, reading, swimming, massage, Qi Gong, art, family\slash friend time, and exercise.
In the article at hand, we limit ourselves to the quantitative analysis and a more high-level analysis of the qualitative survey data and the journal entries. More details on the daily applied practices reported on in those are presented in another publication for reasons of space.


\subsection{RQ1: Changes in Mindfulness Attention Awareness and Daily Perceptions}\label{sec:rq1} 

Research question 1: ``How did participants mindfulness attention awareness and daily perceptions of their experience of life change?'' was split into three sub-questions which we answer in the following.

For the qualitative analysis, Van Dam's suggestion was taken into consideration to analyze the concepts on their theory-based conception~\cite{VANDAM2010805} to better capitalize the benefits of Mindfulness. Given that the concept of Mindfulness per se is multifaceted~\cite{sutton2016measuring}, in order to optimally analyze the effects and experiences in the participants, it was necessary to handle them separately. Hence the authors decided to divide the concept of ``Mindfulness Attention Awareness'' into the concepts of ``Mindfulness'', ``Attention, and ``Awareness''. This division allows to better explain the experience of the individuals for the encompassing concept. At the same time, this separation made it possible to implement one more of Van Dam's suggestions, which is to consider the contribution of traditional Buddhist conceptualizations and psychological implications~\cite{VANDAM2010805}. This was enabled by the long-term meditation study background of the first author and the psychology expertise of the third author.
    
\subsubsection{Does the intervention bring about change in the participants Mindfulness Attention Awareness? (RQ1a) 
}\label{sec:rq1a}

\begin{figure*}
    \centering
    \includegraphics[width=\textwidth]{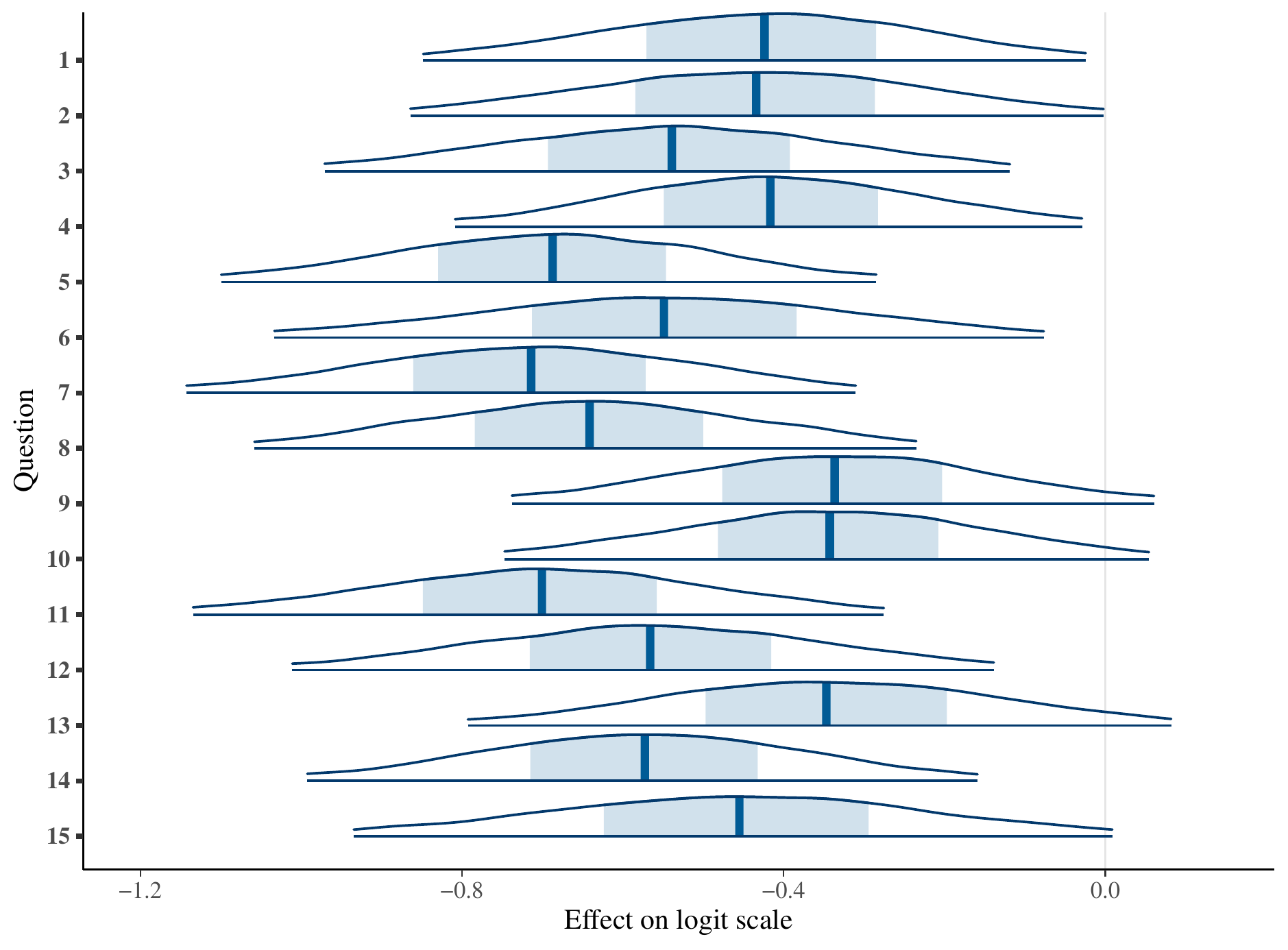}
    \caption{Density plots computed from posterior draws for MAAS. The densities are cut off at 95\%, the blue vertical line is the calculated value of the model for this item, and the shaded area is the 50\% uncertainty interval. We can see a number of questions crossing zero (no effect observed). Most effects are negative (to the left of zero), which means that participants rated more negatively at $t_1$ (exit survey) than at $t_0$ (entry survey), so they were under the perception that their mindfulness attention awareness had decreased.}
    \label{fig:mcmc_MAAS}
\end{figure*}

The MAAS instrument (App.~\ref{app:maas-instr}) consisted of $15$ statements to agree or disagree with. In the \textbf{quantitative findings}, eleven of the ratings indicated a significant difference at $t_0$ vs.\ $t_1$: Q$1$--$8$, $11$--$12$, and $14$. 
\noindent
In all the above cases the effect was negative, i.e., the responses were higher at $t_0$ than at $t_1$, as visible in Fig.~\ref{fig:mcmc_MAAS}. For the other predictors, \textsf{age} and \textsf{gender} did not have a significant effect, while \textsf{occupation} was significant (negative) for Q$2$, i.e., ``I break or spill things because of carelessness, not paying attention, or thinking of something else.''

\begin{figure*}
    \centering
    \includegraphics[width=\textwidth]{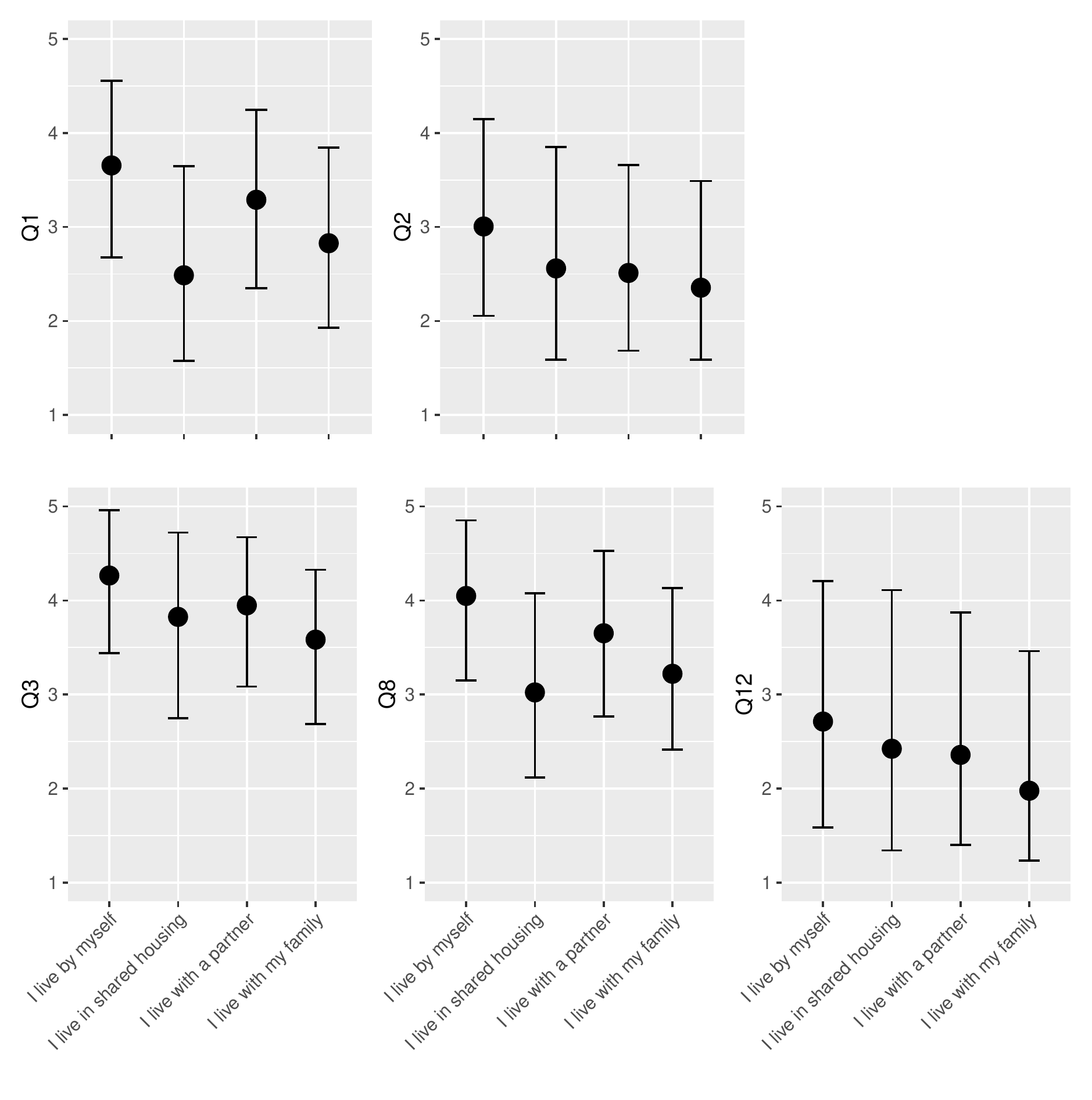}
    \caption{Conditional effects concerning the predictor \textsf{living condition} and its categories for the five MAAS ratings that were significant. By fixing all predictors at their mean or reference levels, a view of each category's effect for our predictor \textsf{living condition} is obtained. The bars correspond to the $95$\% credible interval, while the dot indicates the mean. One or more categories were significant, i.e., from left to right on the horizontal axis in MAAS Q$1$ the 2nd and 3rd, in Q$2$ the 4th, in Q$3$ the 4th, in Q$8$ the 2nd and 4th, and in Q$12$ the 4th category were significant and, hence, the main drivers for the predictor to be significant. The vertical axis shows the Likert scale values.}
    \label{fig:ce-living}
\end{figure*}

Additionally, the predictor \textsf{living condition} was significant (negative) in Q$1$--$3$, $8$, and $12$ (items listed in App.~\ref{app:maas-instr}).
\noindent
Figure~\ref{fig:ce-living} provides an overview of what this implies on the outcome scale, the Likert scale for the five questions, where \textsf{living condition} was significant.
This result could indicate that people who live with their family may be more occupied with the well-being of the ones around them that they feel responsible for, or that they tend to be more preoccupied because they do not find sufficient time and space for themselves to unplug and recharge.

In summary, a number of significant effects were found. Considering the temporal variable, five questions indicated a difference between $t_0$ and $t_1$ (generally speaking subjects answered with higher values at $t_1$).
We also ran the models for the predictors of number of sessions attended as well as number of sessions attended live and number of sessions attended recorded and we see the same overall effect, see App.~\ref{sec:app:det-sess-numbers}.

To make sense of the negative shift in the quantitative results state that the participants rated themselves worse than at the start of the study, we found a large amount of evidence in the qualitative data that shows quite the opposite - that participants have become way more aware. The reason for the more critical self-assessment may well be a consequence of increased awareness, see Fletcher and Bailey~\cite{fletcher2003assessing} for details on issues with self-awareness assessment. 

We move on the \textbf{qualitative results} for this question, answered by the responses coded under theme ``main learnings (changes) identified by the participants'' (see Tab.~\ref{tab:themes}).

\paragraph{Theme: Main learnings (changes) identified by the participants.} Participants described how they experienced the changes in their perception during the course, mainly in awareness, mindfulness, and attention. They commented how these changes influenced their relationship with themselves, with others, and with their environment.

\paragraph{Awareness.} Participants reported significant changes on this construct. The participants mentioned enhanced awareness about themselves, identifying their breathing pattern during the day, for example.

\begin{quote}\footnotesize
    \textit{``I noticed that I am more aware of my full breathing during the day and not only during yoga\slash meditation.''
(participant 31, run 2, journal, Feb 4 2021)}
\end{quote}

The participants' reports focus mainly on the awareness of the body, as the previous quote mentioned and as it can be read in the section below.

\begin{quote}\footnotesize
    \textit{``I will so much like to really feel all my sensations and be aware of my body, for instance, I still have a lot to learn, but I can see the progress over all these weeks, though.''
(participant 14, run 2, journal, Mar 30 2021)}
\end{quote}

In this case, the participant commented that they had noticed progress in the process of awareness of sensations and their own body during the weeks that go from the breathing course.
In addition to noticing changes in their perception of sensations, breathing, and the body, another participant wrote about the changes in their general needs and also actions taken towards those them.

\begin{quote}\textit{\footnotesize
    ``I think this experiment is making me more aware of my needs and doing what is nice for my body and soul.''
(participant 31, run 2, journal, Feb 6 2021)}
\end{quote}

The changes in the daily perceptions of the participants were not limited to the body and the self; instead, they expressed that the changes during the course motivated them to put a higher score for their day-to-day life when writing the journal. The participant below describes in a very clear way the motivation to raise their daily score. 
\begin{quote}\footnotesize
    \textit{``I remember that I started this voting above with a 7, which was something like `yes, the day was ok, I feel ok, nothing unusual happened' [\ldots] I thought that a 5 or 6 is too low for that. But know I raised that up to an 8, because I'm aware of a change that happened in the last weeks.''
(participant 21, run 2, journal, Feb 16 2021)}
\end{quote}

The tools that mainly led the participants to be more aware of their emotions and to being able to control them were analysis and reflection. 
The above quotes express the variety of changes in awareness that the participants experienced during the weeks of the course at different levels and areas.

For \textit{consequences of being present}, the qualitative responses included more intentionality, better listening, better connection with people, better self-care, more relaxed, happier, less obsessive-compulsive, more alertness, higher effectiveness, calmness, softness, more creative, more productive, being relaxed and energized, more joyful, and feeling inspired.

\paragraph{Mindfulness.}
Participants also identified changes in mindfulness. Some were able to recognize them and even link them to other thought processes such as focus and the feeling of happiness.
\begin{quote}\footnotesize
    \textit{``Meditation has helped me greatly, I have been able to focus more on the present moment, have more focus and feel genuine joy more often, and not only thinking that I am enjoying but having a divided heart and mind.''
(participant 14, run 2, journal, Mar 16 2021)}  
\end{quote}

In this quote, it is clear that the tool used is meditation; at the same time, breathing exercises were also mentioned as the way to work with mindfulness. In the same way, concentration and focus were vital for several participants in their mindfulness process. 
\begin{quote}\footnotesize
    \textit{``There are things I have no control on. When these things happen it is hard to concentrate on what I am doing. but reminding what is important at this moment and breathing is helping me recently to get my thought together.''
(participant 53, run 1, journal, Oct 28 2020)}    
\end{quote}

The previous paragraph illustrates how a participant, focused on mindfulness added to the breathing practice, manages to focus and organize their thoughts. The participant can also identify what makes it difficult for them to manage their concentration, which implies a process of awareness. 
\begin{quote}\footnotesize
    \textit{“In reflection, lots of similar experiences in the past were less pleasant, due to my lack of self awareness and poor mindfulness which allowed me to defuse disruptive behaviours triggered by stress (e.g., loosing focus, feeling insecure, cognitive overload, etc.)”.
(participant 82, run 1, journal, Nov 22 2020)}   
\end{quote}

This participant even compared how they lived their past experiences in a different state of awareness and mindfulness and concluded that stress was the reason for those behaviors. They also identified their emotions and behaviors. Likewise, they mention that it was thanks to the reflection that they managed to have this insight after participating in the breathing course.

\paragraph{Attention.}
The analysis also showed how the intervention influenced the attention of the participants. They explained how they experienced these changes in the quotes below.
\begin{quote}\footnotesize
    \textit{``I did tonight the breathwork practice, and I feel emptier and lighter. My mind clearer, fewer thoughts, and more directed. Higher sensuality with my body, higher sensitivity to music.''
(participant 18, run 1, journal, Sept 24 2020)}    
\end{quote}

This participant explains the results of using the breathing techniques. They describe how their senses came into focus and clarified. They also express a feeling of lightness and increased sensitivity and connection.
\begin{quote}\footnotesize
  \textit{``I will say, the breathing exercise guide you hosted really helped with clearing my mind from thinking far ahead and behind. I stayed present all throughout the day. I felt more at peace with myself. [\ldots] Coming back into it with your guidance reminded me again of why it's so important. I am relaxed and willing to take on whatever task comes my way.''
(participant 32, run 1, journal, Sept 25 2020)}  
\end{quote}

This quote describes how the breathing exercises focused the participant's attention on the present, as they commented that they cleared thoughts of the future and the past. In the same way, it served as a tool for enhanced mindfulness and relaxation, motivating the participant to focus their attention on future tasks as mentioned.

The excerpts of the daily journal show that the participants carry out more than one thought process at the same time. Sometimes they are not aware of the combination of these processes. However, when describing their experiences, the combination of attention and, in some cases, also mindfulness is clear. Similarly, most of the participants' insights happened through reflection, which was motivated by writing every day. Keeping a journal pushed participants to rate their day and reflect on what was most important to them. That brings us to RQ1.2, the rating over participants experience over time.

\subsubsection{How did the perceptions of life experience progress over time? (RQ1b)}\label{sec:rq1b}

\begin{figure*}
    \centering
    \includegraphics[scale=0.4]{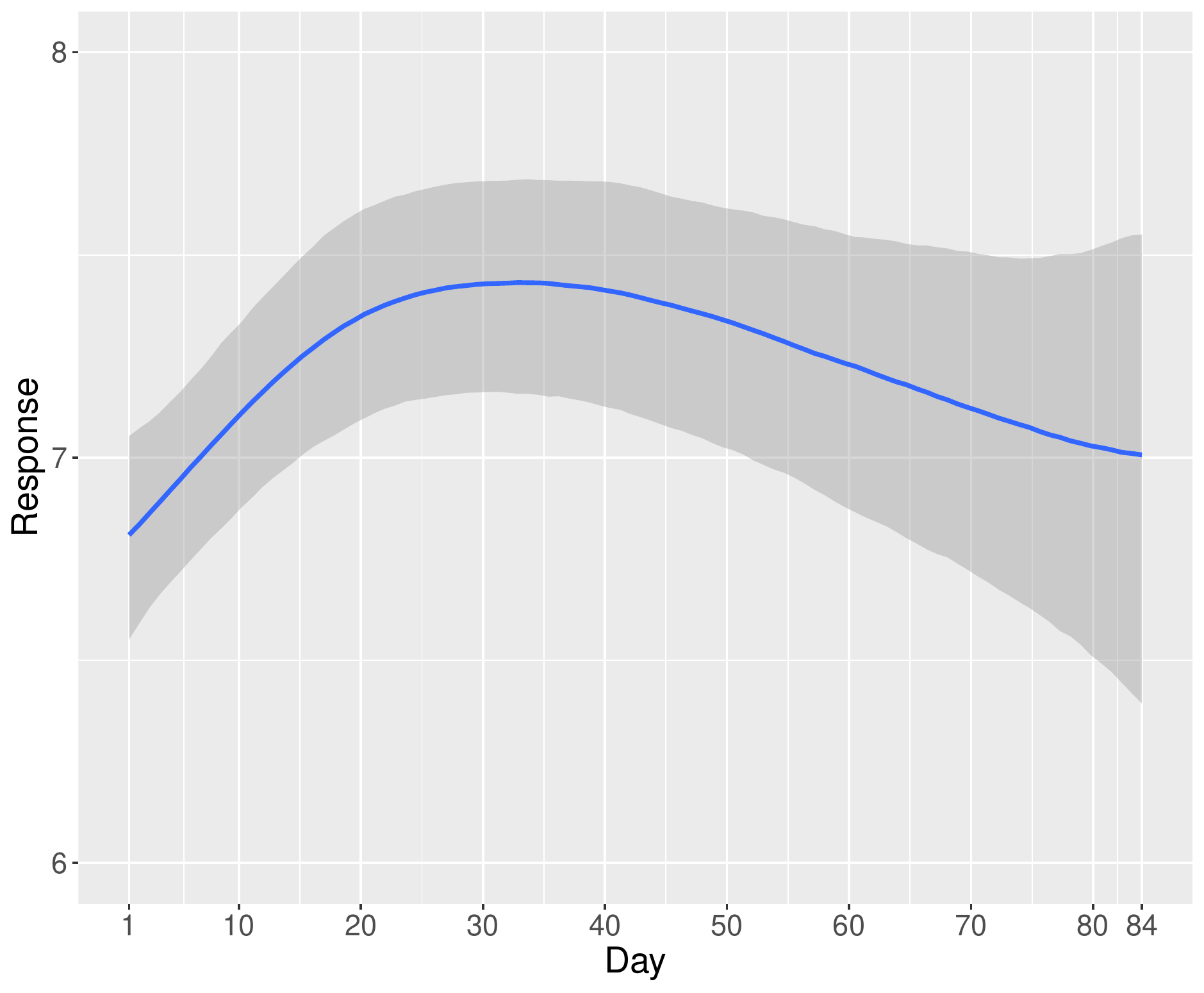}
    \caption{Trend for daily survey ($N=1646$). The blue line indicates the median while the band signifies the 95\% credible interval. On the vertical axis we have the response on Likert scale ($1$--$10$), while the horizontal axis indicates the day. Even though an initial positive trend is visible, due to the uncertainty (especially in the later part of the study) one cannot draw any conclusions.}
    \label{fig:daily}
\end{figure*}

Participants rated their day in the journal entries from $1$ (Really bad) -- $10$ (Absolutely great).
A positive trend is present for two thirds of the intervention, then dips back down towards the end, Figure~\ref{fig:daily} provides a visual overview. Given the uncertainty over time, we cannot claim a significant trend.
While still showing an absolute improvement from beginning to end, the reversed tendency was significant enough to look into. 
We attribute the observed slight decline towards the end to two effects: 1) The newness of the intervention is wearing off and the end of the study is in sight. 2) There is a plateau effect after practicing for a while that shows up as a less positive rating of items. 
Later conversations with participants confirmed these hypotheses.

From the quantitative analysis we see a positive trend that is indicative but not conclusive. However, the qualitative data from the journal entries around the theme ``Integration of techniques in daily life'' support the positive trend as follows.


\paragraph{Attachment and Integration of techniques in daily life---Reflections.}
Participants described feelings, situations, and emotions that, through reflection, they noticed that they lived differently.

\begin{quote}\footnotesize
    \textit{``This daily journal exercise is making me feel really good about little things in the day that I might've otherwise forgotten... Rise 2 Flow is making me reflect!''
(participant 10, run 1, journal, Sept 29 2020)}
\end{quote}

This participant wrote how, through reflection, they were more aware of things going on during the day that, otherwise, they would forget. They also mentioned how to write every day about their day creates a feeling of good.

\begin{quote}\footnotesize
    \textit{``I feel a lot of gratitud to the people and situations I am encountering. I feel more compassionate about other souls and I connect myself quicker than three months ago.''
(participant 18, run 1, journal, Dec 12 2020)}
\end{quote}

This quote describes how the participant felt more compassion, a lot of gratitude, and enhanced their ability to connect to other people. They are aware of when this change started with the course.

\begin{quote}\footnotesize
    \textit{``There are behavioral patterns of mine that I am trying to observe and see how I can make any changes in them. Previously I was so unconscious about them and I used to notice them long after my actions. I did one of them on Thursday and I recognized it right after.''
(participant 53, run 1, journal, Oct 1 2020)}
\end{quote}

These participants wrote about the wish to observe and later change specific behavioral patterns. They mentioned that it was difficult to identify these behaviors when they took place; instead, they realized them later on after participating in the course. This participant expressed that they were able to locate said pattern right after it happened; this is a notable improvement in awareness that can be used in daily life to spot and improve behavior.

The recurring topics in the daily journal were work, stress, and family and friends. Regarding work, the participants identified it as one of the core causes of stress and, on some other occasions, as a measure of productivity. Stress was present in the descriptions mainly as a result of work or illness. Family and friendly relationships were the principal support that participants used to cope with stress.

Throughout the course, the participants explained how the way they handled stress and work changed, improving control and the feelings linked to them. As for family and friends, the changes focused on enjoying and valuing these interactions more.

\textbf{Summary RQ1:} We answer \textit{``How did participants mindfulness attention awareness and daily perceptions of their experience of life change?"} with \textbf{yes, indicating an improvement}.
While the survey instrument shows a negative trend, the qualitative data and the experiences described in the free-text answers of the survey and the journal entries paint a different pictures.

The qualitative analysis shows how participants focused their attention towards appreciation and gratitude, and became more reflective in reporting on situations that did not go so well. We observe a `growing up' tendency in taking responsibility for their own experience of life and in choosing their focus. One participant concluded:
\begin{quote}\textit{\footnotesize
    ``Awareness and consciousness. Time seems to expand as I feel more effective in processing information and seeing connections. I can pinpoint parts of my body I hadn't realise were sending me signal, and my mind becomes more responsive to information and connections. It somehow becomes easy to notice subtleties and details in images, sensations, text, dialogues, etc.'' (participant 82, run 1, exit survey)}
\end{quote}

Or, by another participant, stated more informally:
\begin{quote}\textit{\footnotesize
    ``Generally, things are good. Took a while to get here, but adulting has finally paid off.'' (participant 29, run 1, exit survey)}
\end{quote}

\subsection{RQ2: Changes in Well-being}\label{sec:rq2}

Research Question 2 \textit{``Does the intervention lead to change in the participants well-being?"} was composed by several subquestions listed in the following. Before we dive into answering each one of them, we present the overall qualitative findings on this topic.

\paragraph{Constructs modified during the course.} The changes expressed by the participants span areas of well-being. Participants explain how various aspects of their lives changed throughout the course. One of these aspects is the way they lived their experiences, which were influenced positively. Some participants expressed perceiving the differences compared to before applying the techniques.

\begin{quote}\footnotesize
    \textit{``I had a new presence experience today. Waking home today without any headphones I started to actually listen.  It was a cool experience, and I really felt as though I could shake off the day, reload and come back with new energy.''
(participant 73, run 1, journal, Oct 13 2020)}
\end{quote}
Similarly, this participant talks about how they perceived different an activity as simple as getting up in the morning. They also mentioned how a small change, not wearing headphones, causing a feeling of well-being and sensation of renewal of energy.

Regarding daily activities, the participants commented on how they organized their routine to carry out meditation and breathing practices. Likewise, they expressed the modification of habits, greater reflection, and mindfulness.

\begin{quote}\footnotesize
    \textit{``I am very proud of that daily meditation\slash breathing practice in the morning before turning on the computer. I also changed my breakfast habits and was more present and reflective more. And besides, I'm motivated to take it to 9 someday. :)''
(participant 21, run 2, journal, Feb 16 2021)}
\end{quote}
The above quote shows the motivation and commitment of the participants to work to improve their well-being. Participants also wrote that they feel proud of their actions and set the goal of rating their day with a 9 in the future. 

On the other hand, participants wrote about their emotions, as the quote below explains.

\begin{quote}\footnotesize
   \textit{``By being present, I was able to overcome the dark thoughts that would have otherwise consumed me.''
(participant 52, run 2, journal, Feb 27 2021)}   
\end{quote}

This participant mentioned having better control of their emotions after practicing breathing. It is a recurrent comment among other participants. At the same time, it is linked to better awareness about feelings.

\begin{quote}\footnotesize
    \textit{``I'm getting better in taking time for some well-being practice.''
(participant 21, run 2, journal, Mar 15 2021)} 
\end{quote}
An important point is that participants commit to integrating wellness practices into their routines. This commitment served as the basis for the changes they later expressed regarding well-being and other areas.

In addition to the previous changes, participants talked about how they perceive their thinking. They expressed they feel more positive.

\begin{quote}\footnotesize
    \textit{``I did again the morning meditation, feel really happy to had a more fresh and positive perspective on the week and the days to come.''
(participant 87, run 1, journal, Oct 23 2020)}
\end{quote}

The quote above shows how this participant feels after meditation. Happiness, fresh and positive perspective are the results they describe. 

Gathering what was mentioned by the participants, it can be seen how they gradually perceive the changes during the weeks that the course lasted. They reflected on how their perceptions were modified and explain small experiences that they perceived differently. They better identify emotions and feelings and can deal with them in an improved way. They also mention how their moods changed. They feel calmer, at peace, and more in control of their emotions. 

Similarly, the participants commented they realized that there are many ways to well-being. One participant wrote that they never tried dancing, for example, because they were ``consistently failing on the things that were supposed to work'', but now they are more open to trying different things. The content of the slides also influenced the changes in well-being. An example is ``pick three things that are important to you'' a participant mentioned that by using this practice, they were able to realize the care they put in others but not on themselves. Finally, other reported improvements were to sleep more, less emotionally burdened overall, resting, and better stress management.

\subsubsection{Does  the  intervention  lead  to  change  in  the  participants’ perceptions of positive and negative experiences? If so, how are their experiences affected? (RQ2a)}\label{sec:rq2a} 

The Scale of Positive and Negative Experience (SPANE) instrument consisted of $12$ questions (see App.~\ref{app:spane-pwb-pts-instr}), that asks one general question: 

\begin{quote}\textit{
    Please think about what you have been doing and experiencing during the past four weeks. Then report how much you experienced each of the following feelings, using the scale below.}
\end{quote}

Responses are on a $5$-level Likert scale and there are six categories of questions, each category having two contrasting question (i.e., positive\slash negative, good\slash bad, pleasant\slash unpleasant, happy\slash sad, afraid\slash joyful, and angry\slash contented).

\begin{figure}
    \centering
    \includegraphics[width=\textwidth]{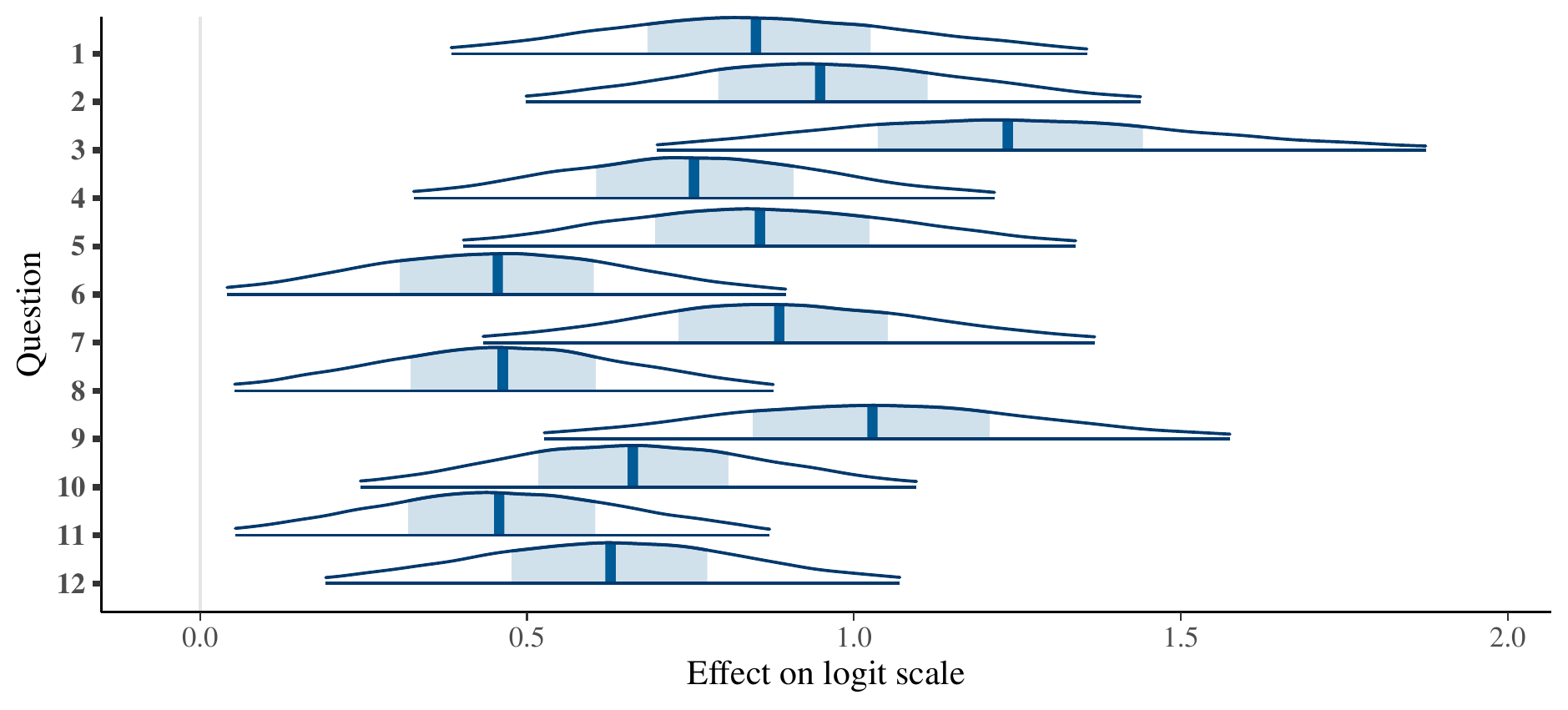}
    \caption{The effects of $t$ for the SPANE instrument. The densities are cut off at 95\% probability mass, the dark blue line indicates the median, and the light blue area is the 50\% probability mass.
    The temporal variable $t$ clearly has an effect (positive) in all questions, i.e., responses were generally speaking higher at $t_1$ (exit instrument).}
    \label{fig:spane-effects}
\end{figure}

All items were \textbf{significant (positive)}, i.e., higher responses at $t_1$, as can be seen in Fig.~\ref{fig:spane-effects}.

The same effects show for attendance of sessions as predictors, i.e. the more sessions a participant attended, the more increase in their SPANE score, see App.~\ref{sec:app:det-sess-numbers}.
Significant effects of the other predictors were the higher the \textsf{age}, the higher the response in Q$9$. Concerning \textsf{gender}, males answered with higher values in Q$3$, Q$6$, and Q$7$.

Both the quantitative data (all rated items) and the qualitative data from the surveys and the journal entries shows a more immediate perception of positive and negative experiences as well as a general tendency towards a more positive perception of participants' daily lives. One participant summed it up as follows:

\begin{quote}\textit{\footnotesize
    ``It's crazy how much some of these modalities/ tools/ phrases\slash meditations\slash breathing exercises can really change a very negative mind!  I'm so grateful that these gifts have come to me in such a timely manner, and that I can hold them to me for the rest of my life.  Now to keep them present and alive, and in everyday use! [\ldots] I have practiced the breathing portions quite often and have a real appreciation for the effect it has on my physical being.'' (participant 23)}
\end{quote}

\subsubsection{Does  the  intervention  lead  to  change  in  the  participants’ psychological well-being? If so, how is it affected? (RQ2b)}\label{sec:rq2b} 

The Psychological Well-Being (PWB) instrument consisted of eight questions (Likert $1$--$7$, see App.~\ref{app:spane-pwb-pts-instr}).
All t parameters are \textbf{significant (positive)}, i.e., higher values at t1, except for Q3, The details are visible in App.~\ref{app:detailedfindings}, Fig.~\ref{fig:pwb-effects} for visualization, correlating with the number of attended sessions (see App.~\ref{sec:app:det-sess-numbers}.
We found significant effects of the predictors for \textsf{Age}, \textsf{Gender}, \textsf{Occupation}, and \textsf{Living conditions}, see also App.~\ref{app:detailedfindings}.

From the quantitative analysis we see that all except one item were rated higher by the end of the intervention, and the qualitative survey data shows participants have had good learning experiences around well-being. For example, one participant reports:

\begin{quote}\textit{\footnotesize
``I am happy that I am aware of my strengths and weaknesses, and that I am able and have learned in my life that we all are meaningful, have a purpose---although that might not be clear or visible most of the time. I am content with my place in life and I have grown to love how my being me makes other people seek my help, presence or comfort.'' (participant 45, run 1, exit survey)}
\end{quote}

\subsubsection{Does  the  intervention  lead  to  change with regard to their positive thinking? If so, how is it affected? (RQ2c)}\label{sec:rq2c}

The Positive Thinking Scale (PTS) (App.~\ref{app:spane-pwb-pts-instr}) consisted of $22$ questions (Yes\slash No answers) and contained some reverse scored items to ensure instrument validity (as noted below). Questions $4$, $9$, $12$, $15$, and $17$--$18$ showed a significant difference between $t_0$ and $t_1$ (Q$9$ and Q$18$ were positive).
The details are provided in App.~\ref{app:detailedfindings}, Fig.~\ref{fig:pts-effects}.

From the quantitative analysis we see that participants did not necessarily change their mind about some parts of their lives or experiences they labeled bad, but there was a shift towards more positivity in a number of items. Overall, participants think more positively at the end of the intervention period.
The qualitative data confirms this, for example:

\begin{quote}\textit{\footnotesize
    Rumination\slash focus on past mistakes is a particular problem for me because I suffer with OCD, but I am trying to get better with dealing with it, e.g., letting thoughts simply pass through (as suggested through Rise2Flow). (participant 10, run 1, exit survey)}
\end{quote}

\begin{quote}\footnotesize
   \textit{I'm a positive person, and I try to embrace this at all levels. I find joy in finding silver linings and in life-long learning. (participant 84, run 1, exit survey)} 
\end{quote}

\begin{quote}\footnotesize
    \textit{How much I've grown and my mindset has brightened since the start of the survey. (participant 42, run 1, exit survey)}
\end{quote}

\subsubsection{How does the well-being fluctuate and vary over the course of the intervention? (RQ2d)}\label{sec:rq2d}

\begin{figure*}
    \centering
    \includegraphics[width=0.8\textwidth]{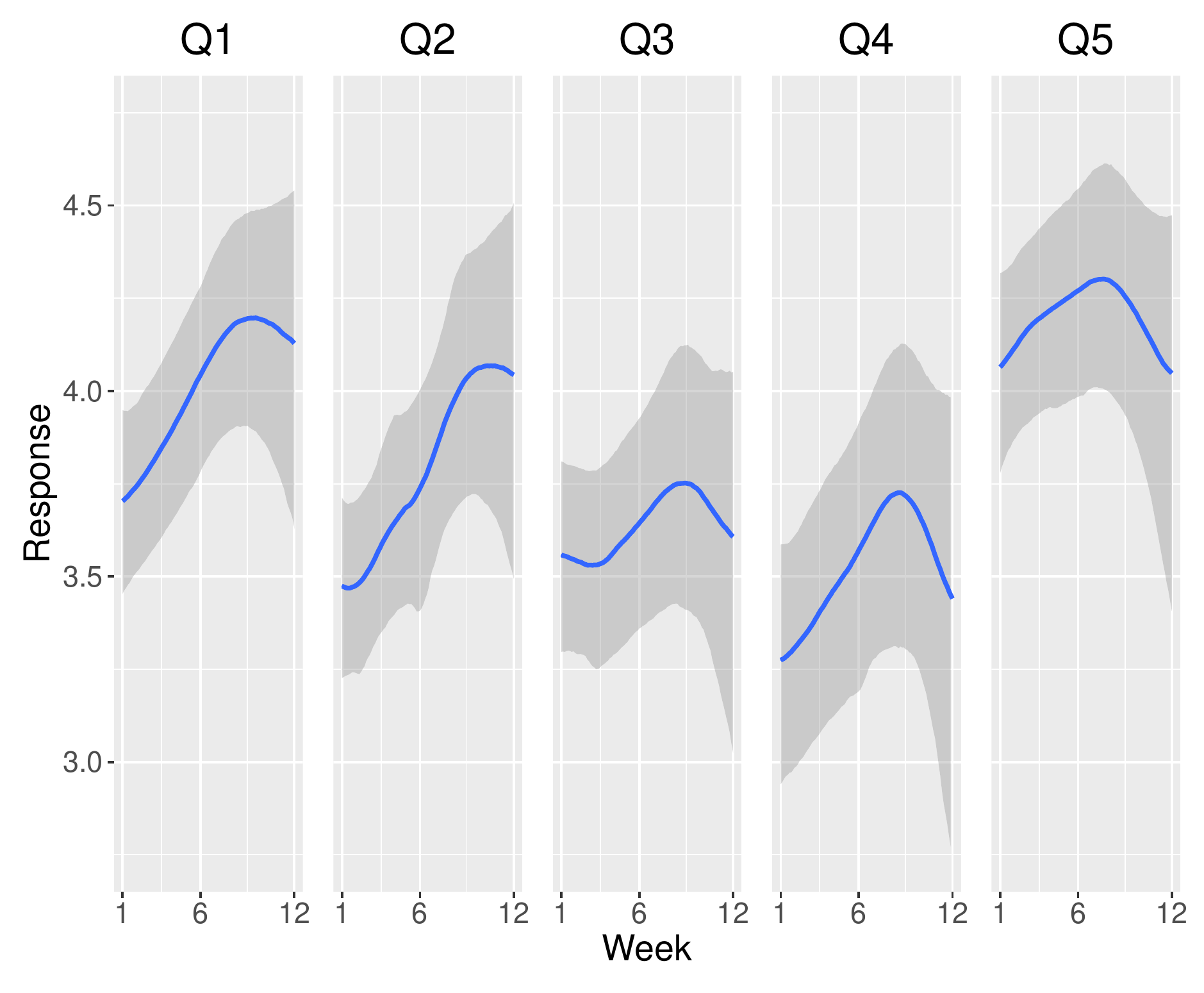}
    \caption{Trends per question for weekly survey ($N=456$). The blue line indicates the median while the band signifies the $95$\% credible interval. On the vertical axis we have the response on Likert scale ($1$--$6$), while the horizontal axis indicates the week. In particular for Q$1$ and Q$2$ one can see a positive trend. Due to the uncertainty no conclusions can be drawn. For Q$3$--Q$5$ this is even more so the case.
    }
    \label{fig:weekly}
\end{figure*}

For the weekly survey, the instrument is shown in App.~\ref{app:who-5-instr}.
Figure~\ref{fig:weekly} provides a visual overview of the daily trends. 
One can see a positive trend for Q$1$, Q$2$ and Q$4$; however, the uncertainty makes it difficult to make any convincing claims. For Q$3$ and Q$5$, this is even more so.

In Fig.~\ref{fig:daily}, one could see the same pattern for the daily rating as in Fig.~\ref{fig:weekly} for the weekly trends for Q$1$, Q$2$ and Q$4$\@. 

From the quantitative analysis, there is an improvement in the ratings of all items of the weekly survey. However, there are trends to be interpreted that raise curiosity.
Houben et al.~\cite{houben2015relation} confirmed that overall, low psychological well-being co-occurs with more variable, unstable, but also more inert emotions.
``Not only how good or bad people feel on average, but also how their feelings fluctuate across time is crucial for psychological health.''~\cite{houben2015relation}
Therefore, it was important to us to look at the development of the WHO well-being scores over the period of the 12 weeks to see the extent and variance of fluctuations.
In that development, we saw a similarly shaped curve in all five items of the weekly survey (see Fig.~\ref{fig:weekly}), with an increase over the first two thirds, then a peak, and then a decrease. Overall, the ratings were still better at the end than at the beginning, yet not necessarily significant. We attribute the decline towards the end to two effects: 1) The newness of the intervention is wearing off and the end of the study is in sight. 2) There is a plateau effect after practicing for a while that shows up as a less positive rating of items.

We observe that there is improvement that shows a positive trend with a specific pattern of fluctuation.
The qualitative data shows progress in some individuals' development of well-being.

\textbf{Summary RQ2:} The research question around participants' well-being is answered with \textbf{yes, an increased well-being} was observable through both quantitative survey items and qualitative data.

\subsection{RQ3: Changes in Perceived Productivity and Self Efficacy}\label{sec:rq3}

Research Question 3 \textit{What are the observable changes in perceived productivity and self-efficacy?} is answered by two subquestions in the following.

\subsubsection{Does the intervention lead to change in the participants perceived productivity? If so, how is it affected? (RQ3a)}\label{sec:rq3a}

The HPQ instrument for Perceived Productivity consisted of eleven questions (with Likert scales varying, going up to $5$, $7$, or $10$, depending on the question, see App.~\ref{app:pp-instr}). 

Only Q$1$ (\textit{How often was your performance higher than most workers on your job?}) shows a significant difference when moving from $t_0$ to $t_1$ (lower responses at $t_1$), which indicates that participants scored themselves lower in performance (which could be due to the continued pandemic working conditions). The details are visible in App.~\ref{app:detailedfindings}, Fig.~\ref{fig:pp-effects}.
This effect confirms the finding indicated by the scores of the MAAS, where participants became more self-aware in general, also noticing when they are off.

Qualitatively, \paragraph{at work} was a recurring theme in the journal entries, especially as a cause of stress in the participants. The perceptions about productivity are what participants used to measure their job performance. The following lines explain how they perceived changes in the area, as mentioned earlier.

\begin{quote}\footnotesize
   \textit{``I have insanely increasing workload at work. But am proud that I am able to stay calm today and focus not only on work but also on other positive things.''
(participant 92, run 2, journal, Feb 2 2021)} 
\end{quote}

The quote above, although brief, describes the positive changes in this participant, better focus, and calmness beyond work.

\begin{quote}\footnotesize
    \textit{``The breathing practice helped me to get a nice state in my mind and reflects in things like today. More awareness made me feel like days are longer but productive.''
(participant 68, run 1, journal, Oct 29 2020)}
\end{quote}

This participant linked awareness and the feeling of being productive even though he feels the days have been longer. 

\begin{quote}\footnotesize
    \textit{``All together, all the awareness of the program has helped me to be more focused and present, enjoying what I am doing, what I don't enjoy, and still finalize it without falling in the temptation of getting distracted by the first thing that pops in my mind (still happens, but is getting better).''
(participant 87, run 1, journal, Oct 23 2020)}
\end{quote}

The quote above explains how this participant has managed better the distractions around and became more focused and present. Besides the change in productivity, they mentioned that they enjoyed more the activities they are doing.

In conclusion, and based on participants' comments, a better awareness resulted in better productivity. The participants seemed to be more relaxed also, better focused, and function better.

\subsubsection{Does the intervention lead to change in the participants’ self-efficacy? If so, how is it affected? (RQ3b)}\label{sec:rq3b}

The Self-Efficacy (SE) instrument (App.~\ref{app:se-instr}) consisted of ten questions (Likert $1$--$4$). Questions $6$, $7$, and $9$ showed a significant effect (positive), i.e., higher responses at $t_1$. 

\begin{itemize}
    \item[Q$6$] I can easily face difficulties because I can always trust my abilities.
    \item[Q$7$] Whatever happens, I'll be fine.
    \item[Q$9$] When a new thing comes to me, I know how to handle it.
\end{itemize}

The details are visible in App.~\ref{app:detailedfindings}, Fig.~\ref{fig:se-effects}. Concerning the other predictors, no significant effects were present, i.e., it is not clear which predictors drove the significant difference between $t_0$ and $t_1$.

Qualitatively, self-efficacy was another area that changed in participants during the course. The following quotes show the perceptions of the participants in their performance in work and daily activities.  
\begin{quote}\footnotesize
    \textit{``I did the most important tasks (rocks) I had set for the day which made me feel great, so much better than the overwhelming feeling I often have when looking through my six page to do list.''
(participant 73, run 2, journal, Feb 2 2021)}
\end{quote}

This participant is using one of the organization techniques presented during the course. They express the positive results of having implemented the rock and sand technique and how this translates into a feeling of well-being compared to similar situations in the past. 

\begin{quote}\footnotesize
    \textit{``Spending a whole day in online meetings but remained quiet and calm, not loosing patience while knowing that I was doing rather duties than something being passionate about.''
(participant 46, run 1, journal, Sept 23 2020)}
\end{quote}

The quote above shows how this participant managed to remain calm and perform their duties even though they expressed they were not passionate about it. It is essential to mention that the activities performed by this participant are mainly online. This scenario may add more stress than usual on this participant; even so, they achieved to keep calm.

\begin{quote}\footnotesize
    \textit{``I've become so much better at looking at what I actually accomplished rather than what I did not prioritize in order to achieve that.''
(participant 73, run 1, journal, Oct 5 2020)}
\end{quote}

This participant has changed the way they address their activities. They decided to focus on what is achieved instead of anything else. With this, the feeling of accomplishment improves, and so the self-efficacy.

\begin{quote}\footnotesize
    \textit{``I tried again the technique of taking five minutes and visualising who I wanted to be in that conference. I focused on trying to convey my excitement about my work, my good humour in connecting with my peers, my excitement to learn new things. I felt very successful in doing so. I noticed how my entire attitude changed and the day did not seem so exhaustive anymore.''
(participant 82, run 1, journal, Oct 25 2020)}
\end{quote}

Learnings from the presentations previous to the breathing session also played a role in the changes in self-efficacy. This participant explains in the quote above how a visualization technique helped to perform better at a conference. They also noticed changes in their attitude that had an impact during the whole day.

\textbf{Summary RQ3.} 
We answer RQ3a, perceived productivity, for now with \textbf{inconclusive}.
The quantitative analysis showed only one item with a significant change over time, which indicated slightly less productivity. 
However, the analysis of the qualitative data showed a lot of examples of how participants had improved their ways of working, scheduling and completing tasks.

We answer RQ3b, self-efficacy, with \textbf{yes, positively}.
From the quantitative analysis, we observe three significant effects, all positive, towards more self efficacy, which is confirmed by the qualitative data in the journal entries.

\section{Discussion}
\label{sec:d}

In this section, we discuss the findings, relate them to other work in the field, and then to a bigger societal picture. We point out limitations as well as various aspects to be taken into account for future work.

\subsection{Significance}
In terms of significance, the following main questions arise: 
\begin{enumerate}
    \item How do the results compare to other modalities and the state of the art? 
    \item What is the impact of an online setting versus in-person?
    \item What is the magnitude of the impact compared to other contextual factors? 
\end{enumerate}

\paragraph{State of the Art Comparison?}
The most well-known program to increase well-being and resilience for IT people is the mindfulness program ``Search Inside Yourself'' by Google~\cite{tan2012search}. It was started in 2007 by Chade Meng Tan. On their advisory board was, amongst others, Jon Kabat-Zinn (work detailed in Sec.~\ref{sec:mindfulness}). A spin-off leadership institute now teaches the program and certifies instructors. Like Rise 2 Flow, they combined a traditional introspective practice (meditation, pranayama) with scientific foundations and self-development topics.
This program is what Bernardez et al.'s work~\cite{bernardez2018experimental} is based upon and evaluated with a software engineering population. Half of their student population practiced mindfulness during 6 weeks, and then the complete population participated in an experiment evaluating the efficiency and effectiveness in conceptual UML modeling. The effectiveness of the program in comparison to ours cannot be established as we did not carry out a UML modeling experiment, but evaluated broader concepts on established psychological scales.
\\
As the breathwork technique of our study has not been evaluated empirically before, we do not have data to contrast software engineers against a more general population. 
\\
We are aware of only one study that has compared a similar type of breathwork to other modalities~\cite{seppala2020promoting}, and it showed the greatest impact benefitting six outcomes (depression, stress, mental health, mindfulness, positive affect, and social connectedness). 
We see a similar outcome in our study in a decrease in stress and an increase in positive affect across the data, and an increase in mindfulness in the qualitative data.
No study has yet compared the breathwork technique that we used in our study, so a more detailed comparison might bring insight in form of first data points for the usefulness of breathwork in general in comparison to other de-stress modalities.

\paragraph{Limited Impact in Online Setting.} Interestingly, it was the lock-downs in many countries and subsequent mental and emotional challenges experienced by students, colleagues, family and friends that motivated us to carry out this study in the first place. The online setting was chosen such that we were able to offer a relieving intervention during the restrictions.
A traditional setting would be a designated physical location where participants meet once a week in a safe space that is specifically prepared for an undisturbed session without distracting technology or disruptions from outside. Such a setting allows for an immersive experience on a different level and, often, much deeper transformation and restoration. During Rise 2 Flow, participants dealt with network service outages, software updates, and usually practiced in their living room turned make-shift office turned trying-to-be attention-restorative environment. On one hand, this certainly limited the benefits that could be received through this modality and, on the other hand, it opened the intervention to a much wider group of participants from around the world.

\paragraph{Magnitude of Impact}
Software developers create the most complex systems in the world, and need a high attention capacity for intellectually taxing tasks in often distributed team constellations. Furthermore, they need empathy for collaborators and clients alike and usually work under time pressure. This study looks into the probably most accessible mechanism to regulate and restore the nervous system - breathing. While the study focuses on a specific framing as intervention (the whole \textbf{program} with the breathing plus reflection practice for specific topics), the \textbf{breathing technique by itself} can also be used individually for a few minutes of reset in any situation. Participants are trained on a technique that helps both to increase resilience by building long-term capacity (see results for well-being Sec.~\ref{sec:rq2} and self-efficacy Sec.~\ref{sec:rq3b}) as well as short-term recovery (see quotes from journal entries, e.g. p.~\pageref{quote:panic}).\\
An influence of season and weather may be present, but cannot be analyzed to a meaningful degree with the collected data as we had participants from all around the globe.
\\
The restrictions in our daily lives due to the Covid may have led to a higher stress level for many participants, which was reflected in many survey answers relating to working from home with implications of either loneliness or taxing family situations. Despite this effect, it also showed the intervention was timely and useful, as mentioned by these participants: 
\begin{quote}\footnotesize
    \textit{
    This rating makes me think about where I'm at in life and how I view myself within my surroundings and social community. Reflecting on social relationships gave me pause because they are supportive and kind, but at times I feel so alone. This is largely because of the pandemic and stay-at-home living. I am still grateful for them and can accept the tragedies as well as the beauties of these new living circumstances. I am also realizing that I can be more engaged in the activities that mean a lot to me; it's easy for me to detach when I feel overwhelmed. I'm reminded of the importance of prioritizing rather than letting everything go.} (participant 42, run 1, exit survey)
\end{quote} 
\begin{quote}\footnotesize
    \textit{It's interesting to do this exercise during a global pandemic in California\ldots We are living in such difficult times in so many ways right now. Interesting to reflect on whether the past was `good' or `bad', considering the current situation. Like, in comparison, the past seems like it should have been so much joy all the time---but of course we took being close together in groups and hugs for granted back then. This makes me sad. I do find myself savoring and really appreciating time with friends and family more now than in the past, even though times are hard right now.} (participant 67, run 1, exit survey)
\end{quote}

\subsection{Observations and Implications}

\paragraph{Policy.} There are implications for policy and resulting applications of well-being indicators, as society becomes more aware of the importance of mental health. For in-depth discussion, see Pavot and Diener~\cite{pavot2004subjective} who call for a national well-being index to inform policy makers specifically in the area of aging, as a first step towards a happiness index like already established in other countries. While advertising for the study, we observed a general acknowledgement of the importance of supporting well-being. At the same time, it seemed that often the support did not go beyond the acknowledgement. Mental and emotional health can only improve if individuals, organizations and institutions alike take responsibility.

\paragraph{Psychographics versus Demographics.} Before the study, we had been wondering whether personality traits would show a difference in how participants benefit from the practice and how their awareness shifted. Therefore, in the exit survey, we added a Mini IPIP personality test\footnote{While we are aware that IPIP is by now considered controversial in terms of its statistical validity, to this date it is still the most widely and commonly used personality test.}, so we could control for personality in the results. However, IPIP did not reveal anything in the analysis. This does not mean that personality has no influence; it only means we did not see conclusive evidence for a particular personality trait as observable via Mini IPIP\@. 

\paragraph{Role models.}
As senior academics, we are role models---whether we want it or not---simply because we speak in front of students, we teach and supervise. If we do not model taking care of our nervous systems, including physical, mental and emotional health, we not only neglect ourselves, but also fail to provide our students with guidance. We are not advocating for every senior academic to give lectures on the topic. Instead, we advocate for every single person prioritizing their well-being over external demands so we can operate from a well-resourced place and thereby deliver better service to the world. Living into that as a role model is a more effective way of teaching it than postulating the theory.
Our participants commented on this as well:
\begin{quote}\footnotesize
    \textit{``It was cool to see that profs were participating in that study as well."}\\ (informal conversation with participants 77 and 78 after Rise 2 Flow 2)
\end{quote}

\subsection{Limitations and threats to validity}

\paragraph{Sampling Bias.} The participants for both runs of the experiment were recruited across a range of personal and online networks, including the global personal network of the first author, university networks, mailing lists, online spaces, and social media channels. While it is not a classical convenience sample because of the number of channels used for broadcasting, it can be seen as an extension thereof~\cite{baltes2020sampling}. However, all these networks are initially based on connections to the first author, which introduces a potential sampling bias. We mitigated this threat to the best of our ability by requesting re-posts and further distribution of the call for participation in the several hundred emails and posts the first author put out for recruitment across the disclosed variety of channels. We follow the reporting guidelines proposed in~\cite{baltes2020sampling}.

\paragraph{Self-selection Bias.} It is possible that people who are drawn to participate in a study like ours are not a representative sub-population of the overall study population. By repeating the experiment and learning more about the participants through the surveys and follow-up interviews we aim to learn more about that aspect. 

\paragraph{Response Bias.} We used standard validated instruments in our survey that prevent response bias to the degree possible, following recommendations by Dillman et al.~\cite{dillman2014internet} by, for example, putting content questions before demographics. 
A remaining response bias is due to the fact that the participants got to know the first author as instructor, which may have introduced a bias in their free-text responses. We mitigated this by letting participants know, during sign-up, that their data was going to be anonymized before analysis.

\paragraph{Construct Validity.} We used validated scales as referenced. We focus on the breathing element of the intervention, but there were other modalities offered for the days in between if participants wanted more, namely guided meditations and journal reflection prompts.

\paragraph{Measurement Validity.} There might be a threat to measurement validity by participants getting tired while answering the $78$ items on the entry\slash exit survey. Our pretests showed that the survey could be completed within ten minutes.

\paragraph{Internal Validity.} There is a threat to internal validity as we had sessions that combined reflective conversation in the larger group with the breathing practice.
As it is important for continuance of a course to build a relationship with the participants, there is no way to differentiate how much the community aspect may have contributed to the positive effects of the breathing exercises. In statistical language, given data, it is hard to control for a `community effect'.\\
There is a threat to validity in terms of the effects of the breathwork in comparison to the effects of the topic conversations. The awareness raising is happening on a neurological/unconscious level by breathwork, and on a mental/rational level by topic presentations. Most participants liked both, and a few preferred only one of the components - for those participants there could be a stronger influence from one of the program components over the other. To the best of our knowledge (from being an instructor for years as well as from the qualitative data) we see that the main changes are coming from the breathwork for most people, and that the topics do have an impact that is minor in comparison. \\
A few participants reported later on to have performed breathwork in additional practice, which may have influenced their overall results. To the best of our knowledge, there were only few of them that did additional practice (e.g. for a few minutes before falling asleep after a stressful day), so the threat to validity is minimal.\\
We did not have a control group for several reasons. We tried establishing a control group in the first pilot (see Sec.~\ref{sec:interv}), and got zero responses. It can be done with a waitlist approach where future participants serve as comparison group (as done by Bernardez et al.~\cite{bernardez2018experimental}), but that introduces a number of biases as well, for example that respondents are self-selected and in favor of trying the approach as opposed to a random control group, and that they are primed for the surveys by the time they participate in the intervention. Consequently, it is controversial (also in medical studies) whether this is a good approach. If we were to select a random control group, there would be no `placebo' to mask whether participants receive the actual intervention under research or a different one. Bernardez et al.~\cite{bernardez2014controlled} did that in a one-session experiment to test whether cognition and concentration increase after a meditation session compared to a session about how to give good presentations, but this approach is less feasible in a 12 week intervention. A non-equivalent control group post-test-only design~\cite{krishnan2021review} was not feasible either.
Given the drawbacks of the ways of how to work with control groups in this case, we do not think this would strongly increase the confidence in the results.

\paragraph{Conclusion Validity.} The threat to conclusion validity brought about by potential researcher bias is mitigated by correlating insights from quantitative and qualitative data. External validity and generalizability are limited to the demographics of the participant population. Reliability is provided by using validated instruments and standard methods as well as a replication package.

The qualitative data we report on reads positive. That raises the question of whether the thematic analysis was carried out in a balanced manner. We do not have any reports on negative or unwanted effects. 
Most likely, people who did not have the desired effects stopped reporting - and we cannot conclude on causality either way.
We cannot report on evidence we do not have, therefore we are open about analysing what was there and that people who did not feel desired benefits yet dropped out along with the people who had urgent other matters come up (as quoted in Sec.~\ref{sec:demographics}).





\subsection{Relating Back to Theory in Psychology}

This study aimed to research the effects of breathing practice on the mindfulness attention awareness, well-being, self-efficacy and perceived productivity of computer workers. 
The results reported positive changes in these areas. Participants expressed after the breathing workshop to feel more relaxed, calmer, and more in contact with their emotions and, in some cases, with other people. Some of them also mentioned noticing improvement in areas such as creativity and their performance at work.

According to Beck's theory~\cite{hofmann2013science}, the influence of breathing exercises on the participants results in a change in mood and, therefore, in a behavior change. This change was manifested by themselves when commenting that they can better manage their emotions, identify their thoughts and be more present at the moment (Sec.~\ref{sec:rq1}/\ref{sec:rq2}).

Fisher~\cite{fisher} mentions that the state of well-being is achieved by integrating seven basic skills into daily life.\footnote{
\begin{enumerate}
    \item Engage in sustained, constructive, self-controlled goal-directed activity within complex social environments; 
    \item Respond constructively to social challenges; 
    \item Engage in self-controlled, creative, goal-directed activity; 
    \item Engage in and enjoy positive, reciprocal social relationships; 
    \item Engage in present-focused activities of a sensory, meditative, creative, playful or aesthetic nature; 
    \item Achieve a balance between the demands of socially engaged, goal-directed activity and other kinds of activity; and finally 
    \item Understand the nature of wellbeing and the social and environmental conditions required to attain it~\cite{fisher}.
\end{enumerate}
} 

He suggests that a uniform balance between them creates the optimal conditions for well-being, and breathing practices support all of these.
This theory emphasizes that the exercise of the skills that form well-being occurs when the individual interacts with their environment in social relations. Stress plays an essential role in these interactions; individuals will bring their moods and feelings to these interactions, qualifying their own experiences and those of others.

Considering what both theories mention and the results observed in the participants, breathing techniques contribute to the development and maintenance of personal and partially social well-being.

\subsection{Future Work}

We are planning several follow-up studies, namely a larger cohort in a company and a simplified version with less weekly time investment.

\paragraph{Larger Cohort.} We are currently looking for an industrial collaboration partner who can contribute with a larger sample from within one company, so we can see the distribution of effects in a similar work environment. Providing the study in a closed program advertised for in the company and supported by the company would increase the likelihood of employees sticking with the program instead of dropping off.
For that study, instead of the original PWB, we consider using Dagenais'~\cite{dagenais2012psychological} version of the instrument in a study with larger samples from a single company or similar companies as opposed to the wide range of work contexts of the subjects in the study at hand, as they suggest it may be useful to adapt the instrument to make it more specific to evaluate a well-being in a work setting~\cite{dagenais2012psychological}. Their PWB seems to have a strong eudaimonic connotation from the survey participants' point of view.

\paragraph{Simplified Version.} There was feedback on the level of involvedness required for the study, e.g., the $90$ minute live session plus several surveys, so we are considering a reduced version. The questions to solve here are (1) how much can we slim it down with it still being a meaningful intervention, and (2) how much can we simplify the data collection but still get meaningful data.
Most clinical studies get their participants from therapy interventions, and therefore have large numbers and wait-listed control groups because those people are in sufficient mental and\slash or emotional pain to act on it. However, if we offer to intervene before that pain becomes too dire, the intrinsic motivation may also be lower. If the reader ever dropped off a well-being practice after things got a little better, they can relate.

\paragraph{Further Instruments} We are additionally interested in exploring the scales for mystical experiences used by John Hopkins hospital~\cite{barrett2015validation}, as several participants indicated mystical experiences during the sessions. 

Furthermore, as self-connection is the main foundation for relation to others and effective information flow (specifically important for software engineers), especially during Covid~\cite{lucas2021mindful}, we are also interested in working with the self-connection scale by~\cite{barrett2015validation}.

\section{Conclusion}
\label{sec:c}

In this article, we presented the results of an intervention with live group breathing practice to deepen the participants' connection to themselves, framed with a weekly self-development topic. Awareness raising is happening on a neurological/unconscious level by breathwork, and on a mental/rational level by the topic presentations and reflecting upon them in group conversation as well as in personal practice with proposed tools. 
The quantitative and qualitative results indicate that this intervention may be helpful in improving participants' mindfulness attention awareness, well-being, and self efficacy. 
\\
There is a wide selection of wellness classes available outside of work for the person looking, while at work there may be a few generic offerings 
that work on a content level, but often not on a neurophysiological or embodied level.\\
Software engineers have a strong background in rational thinking and work with empirical evidence, so there is a need for programs with adequate language such that software engineers who feel overwhelmed are attracted - science-based and in a safe space, brought to them by someone who can relate to their specific work experiences. This may help sway hesitant software engineers to try out a relaxation and recovery technique, benefitting their personal resilience and well-being and, in turn, their work performance and job satisfaction (important for retention).
Consequently, we see three ways of potential impact by our study: 1) to inform and raise awareness in the research community as well as in practice, 2) to train further cohorts of software engineers and software engineering researchers and educators in restorative practices, 3) to develop tailored programs for companies and higher education that teach these techniques and frame them science-based while still focusing on the embodiment component to increase self-connection.
\\
The main challenge that remains is that the pace of work life is artificially high because of a perceived need for constant competition (e.g. time to market, to offer better service, to increase our skills, etc.) as remarked by several participants in our study to the point where they felt they didn't ``have time'' for restorative practices. The speeding-up of life we have been witnessing over the past decades has consequences for health. In a certain pattern, physical stress is healthy and makes sure that we get certain things done - and those phases of stress needs to be taking turns with phases of recovery (beyond sleeping 6 hours per night). When recovery is not sufficiently given, stress wears on our physical (adrenal fatigue), mental (burn-out), and emotional health (depression and anxiety).
Restorative practices can help us recover more quickly and become more resilient - they do not change the underlying systemic misalignments.
\\
Our vision is that restorative and contemplative practices can support us in recovering a stronger connection to self, such that we have the mental and emotional capacity to reflect on our values and how we live into them. We get to decide every day how we want to continue, and the constraints can be shifted, some immediately, some over time. There are systems with unhealthy dynamics in place, yes, and we can change them -- because we humans are the ones that created them. If we don't like the constant stress and time pressure, let's change the systems and societal structures that create them. Part of that is acknowledging the tendency of the human mind to always want more (and we see how it plays out in our economy), and developing our own practice to stay present with that~\cite{dass1971here}.
The first step towards that from the perspective of our research is: 
\textbf{Let's normalise taking care of our nervous systems as much as brushing our teeth, and thereby improve our physical, mental, and emotional health.} There could be a start of every meeting with a deep breath to become present, someone teaching peers an emergency breathing technique to relax and focus before a presentation, there a well-being course that teaches breathing practices (or other restorative techniques) twice a year at a company, a weekly meditation group that provides community support in addition to daily personal practice (when it comes to personal practice, 5 minutes is always better than nothing). The options are many, the prioritization is an individual choice.

We leave you with a quote from a journal entry that sums up results reflected for a number of participants and that seem worth acknowledging: 
\begin{quote}
    \footnotesize{\textit{
Today was the last day of the 12 weeks. 
I took away a whole new world, that I am still trying to reconcile with. (...) 
Anyway, learnings: be conscious where you put your attention, and hence your energy, what the wonder precious moment is, that I am not different---I am unique, to put intentions to things, how meditation with a intention\slash visualization can change your day, that breathing can 
``make you float'' and have psychedelic experiences, the forgotten joy of dancing, the power of gratefulness and that I am grateful for the bad stuff that happened to me (!), how important it is to love and be kind to oneself, to surrender to feelings rather than pushing them away, the power of small routines (as well as the difficulty of keeping them), that I am not my thoughts or my emotions (what the f*+@?!?!), (...) 
What else can I say, really? THANK YOU!!!} - participant 75, run 1, journal, Dec 10 2020}
\end{quote}

\section{Data Availability}
\label{sec:da}
To support open science, the replication package including the raw quantitative data is available on Zenodo  \url{https://zenodo.org/record/5082388}, which links to a Github repository \url{https://github.com/torkar/rise2flow}. 

The qualitative responses are not available as many of them reveal very personal experiences, deep emotions, and individual life circumstances that might involuntarily disclose identifiable information.

\begin{acknowledgements}
We thank the participants of Rise 2 Flow 1 and 2 for their trust in us to support them in cultivating a personal practice for increased well-being, for their dedication, and for their generous feedback.
The first author thanks Robert Feldt for a helpful discussion of available survey instruments during the design phase of this study, and Sabine and Fritz Penzenstadler for helpful input in conversation and action.
We thank Francisco Gomes de Oliveira Neto and Leticia Duboc for thoughtful feedback on earlier versions of this manuscript. We thank the anonymous reviewers who gave very thorough and thoughtful feedback on an earlier version (shout-out to especially Reviewer 1). We appreciate you.

The computations were enabled by resources provided by the Swedish National Infrastructure for Computing (SNIC), partially funded by the Swedish Research Council through grant agreement no.\ 2018-05973. Part of this research is financed by the Area of Advance ICT at Chalmers University of Technology under no.\ C-2019-0299.
\end{acknowledgements}

%
\section*{Conflict of interest}
The authors declare that they have no conflict of interest.

\bibliographystyle{spmpsci}      
\bibliography{bib}   

%
%
\newpage
\appendix

\section{Survey Instruments}\label{app:instr}
\subsection{The MAAS instrument}\label{app:maas-instr}

The Mindfulness Attention Awareness Scale (MAAS) is replicated from~\cite{brown2003benefits}.

Instruction MAAS: 
\begin{quote}
    Below is a collection of statements about your everyday experience. Using the 1 (almost never) - 6 (almost always) scale below, please indicate how frequently or infrequently you currently have each experience. Please answer according to what *really reflects* your experience rather than what you think your experience should be. Please treat each item separately from every other item. 
\end{quote}

\begin{table}[h!]
    \centering
    \begin{tabular}{p{7cm}|c|c|c|c|c|c}
         & 1 & 2 & 3 & 4 & 5 & 6 \\\hline\hline
        I could be experiencing some emotion and not be conscious of it until some time later. & & & & & & \\\hline
        I break or spill things because of carelessness, not paying attention, or thinking of something else. & & & & & & \\\hline
        I find it difficult to stay focused on what's happening in the present. & & & & & & \\\hline
        I tend to walk quickly to get where I'm going without paying attention to what I experience along the way. & & & & & & \\\hline
        I tend not to notice feelings of physical tension or discomfort until they really grab my attention.  & & & & & & \\\hline
        I forget a person's name almost as soon as I've been told it for the first time.  & & & & & & \\\hline
        It seems I am ``running on automatic,'' without much awareness of what I'm doing. & & & & & & \\\hline
        I rush through activities without being really attentive to them. & & & & & & \\\hline
        I get so focused on the goal I want to achieve that I lose touch with what I'm doing right now to get there. & & & & & & \\\hline
        I do jobs or tasks automatically, without being aware of what I'm doing. & & & & & & \\\hline
        I find myself listening to someone with one ear, doing something else at the same time. & & & & & & \\\hline
        I drive places on `automatic pilot' and then wonder why I went there. & & & & & & \\\hline
        I find myself preoccupied with the future or the past. & & & & & & \\\hline
        I find myself doing things without paying attention. & & & & & & \\\hline
        I snack without being aware that I'm eating. & & & & & & \\\hline
    \end{tabular}
    \caption{The Mindfulness Attention Awareness Scale (MAAS)~\cite{brown2003benefits}}
    \label{tab:maas-instr}
\end{table}

\subsection{The instruments SPANE, PWB, and PTS}\label{app:spane-pwb-pts-instr}

Diener et al.~\cite{diener2009new} proposed a set of related instruments in `New measures of well-being' that includes the Scale of Positive And Negative Experience (SPANE), the scale of Psychological Well-being (PWB), and the scale of Positive Thinking (PTS).

Instruction SPANE: 
\begin{quote}
Please think about what you have been doing and experiencing during the past four weeks. Then report how much you experienced each of the following feelings, using the scale below. For each item, select a number from 1 (Very rarely or never) to 5 (Very often or always).
\end{quote}

\begin{table}[h!]
    \centering
    \begin{tabular}{p{8cm}|c|c|c|c|c}
         & 1 & 2 & 3 & 4 & 5  \\\hline\hline
        Positive & & & & & \\\hline
        Negative & & & & & \\\hline
        Good & & & & & \\\hline
        Bad & & & & & \\\hline
        Pleasant & & & & & \\\hline
        Unpleasant & & & & & \\\hline
        Happy & & & & & \\\hline
        Sad & & & & & \\\hline
        Afraid & & & & & \\\hline
        Joyful & & & & & \\\hline
        Angry & & & & & \\\hline
        Contented & & & & & \\\hline
    \end{tabular}
    \caption{The Scale of Positive and Negative Experiences (SPANE)~\cite{diener2009new}}
    \label{tab:spane-instr}
\end{table}

Instruction PWB: 
\begin{quote}
Below are 8 statements with which you may agree or disagree. Using the 1 (Strongly disagree) – 7 (Strongly agree) scale below, indicate your agreement with each item by indicating that response for each statement.
\end{quote}

\begin{table}[h!]
    \centering
    \begin{tabular}{p{7cm}|c|c|c|c|c|c|c}
         & 1 & 2 & 3 & 4 & 5 & 6 & 7 \\\hline\hline
        I lead a purposeful and meaningful life. & & & & & & & \\\hline
        My social relationships are supportive and rewarding. & & & & & & & \\\hline
        I am engaged and interested in my daily activities & & & & & & & \\\hline
        I actively contribute to the happiness and well-being of others & & & & & & & \\\hline
        I am competent and capable in the activities that are important to me & & & & & & & \\\hline
        I am a good person and live a good life & & & & & & & \\\hline
        I am optimistic about my future & & & & & & & \\\hline
        People respect me & & & & & & & \\\hline
    \end{tabular}
    \caption{The Psychological Well-Being (PWB)~\cite{diener2009new}}
    \label{tab:pwb-instr}
\end{table}

Instruction PTS: 
\begin{quote}
The following items are to be answered ``Yes'' or ``No.'' Write an answer next to each item to indicate your response.
\end{quote}

\begin{table}[h!]
    \centering
    \begin{tabular}{p{9cm}|c|c}
         & Yes & No \\\hline \hline
        I see my community as a place full of problems. & & \\\hline
        I see much beauty around me. & & \\\hline
        I see the good in most people. & & \\\hline
        When I think of myself, I think of many shortcomings. & & \\\hline
        I think of myself as a person with many strengths. & & \\\hline
        I am optimistic about my future. & & \\\hline
        When somebody does something for me, I usually wonder if they have an ulterior motive. & & \\\hline
        When something bad happens, I often see a “silver lining,” something good in the bad event. & & \\\hline
        I sometimes think about how fortunate I have been in life. & & \\\hline
        When good things happen, I wonder if they might have been even better. & & \\\hline
        I frequently compare myself to others. & & \\\hline
        I think frequently about opportunities that I missed. & & \\\hline
        When I think of the past, the happy times are most salient to me. & & \\\hline
        I savor memories of pleasant past times. & & \\\hline
        I regret many things from my past. & & \\\hline
        When I see others prosper, even strangers, I am happy for them. & & \\\hline
        When I think of the past, for some reason the bad things stand out. & & \\\hline
        I know the world has problems, but it seems like a wonderful place anyway. & & \\\hline
        When something bad happens, I ruminate on it for a long time. & & \\\hline
        When good things happen, I wonder if they will soon turn sour. & & \\\hline
        When I see others prosper, it makes me feel bad about myself. & & \\\hline
        I believe in the good qualities of other people. & & \\\hline
    \end{tabular}
    \caption{The Positive Thinking Scale}
    \label{tab:pts-instr}
\end{table}

\newpage
\subsection{Self Efficacy}\label{app:se-instr}

The instrument was developed by Jerusalem et al.~\cite{jerusalem1999skala} and based on Bandura et al.'s~\cite{bandura1999self} self-efficacy model.
It is used to assess the individual stress resilience of the participants and encompasses ten items that offer a positively phrased statement on change, challenges or unexpected circumstances which the participant has to rate as ``Not true'' (1), ``Hardly true'' (2), ``Rather true'' (3) or ``Exactly true'' (4).

Instruction:
\begin{quote}
    Please rate the following statements on the basis of the given scale and tick as appropriate:
\end{quote}

\begin{table}[h!]
    \centering
    \begin{tabular}{p{7cm}|c|c|c|c}
         & 1 & 2 & 3 & 4 \\\hline \hline
        When problems arise, I find ways to carry through. & & & & \\\hline
        I always succeed in solving difficult problems, if I try. & & & & \\\hline
        It does not give me any difficulty to realize my intentions and goals. & & & & \\\hline
        In unexpected situations I always know how to behave. & & & & \\\hline
        Even with surprising events, I believe that I can handle them well. & & & & \\\hline
        I can easily face difficulties because I can always trust my abilities. & & & & \\\hline
        Whatever happens, I'll be fine. & & & & \\\hline
        For every problem I can find a solution. & & & & \\\hline
        When a new thing comes to me, I know how to handle it. & & & & \\\hline
        If a problem arises, I can do it on my own. & & & & \\\hline
    \end{tabular}
    \caption{Self efficacy instrument by Jerusalem et al.~\cite{jerusalem1999skala}}
    \label{tab:se-instr}
\end{table}

\newpage
\subsection{Perceived Productivity}\label{app:pp-instr}

The HPQ\footnote{http://www.hcp.med.harvard.edu/hpq} measures perceived productivity in two ways: First, it uses an eight-item scale (summative, multiple reversed indicators), that assesses overall and relative performance, and second, it uses an eleven-point list of general ratings of participants' own performance as well as typical performance of similar workers. 

Instructions PP: 
\begin{quote}
    The next questions are about the time you spent during your hours at work in the past 4 weeks (28 days). Select the one response for each question that comes closest to your experience from ``None of the time'' (1) to ``All of the time'' (5).
\end{quote}

\begin{table}[h!]
    \centering
    \begin{tabular}{p{8cm}|c|c|c|c|c}
         & 1 & 2 & 3 & 4 & 5  \\\hline\hline
        How often was your performance higher than most workers on your job? & & & & & \\\hline
        How often was your performance lower than most workers on your job? & & & & & \\\hline
        How often did you do no work at times when you were supposed to be working? & & & & & \\\hline
        How often did you find yourself not working as carefully as you should? & & & & & \\\hline
        How often was the quality of your work lower than it should have been? & & & & & \\\hline
        How often did you not concentrate enough on your work? & & & & & \\\hline
        How often did health problems limit the kind or amount of work you could do? & & & & & \\\hline
    \end{tabular}
    \caption{Perceived Productivity from the HPQ}
    \label{tab:pp-instr}
\end{table}

\begin{itemize}
    \item On a scale from 0 to 10 where 0 is the worst job performance anyone could have at your job and 10 is the performance of a top worker, how would you rate the usual performance of most workers in a job similar to yours?
    \item Using the same 0-to-10 scale, how would you rate your usual job performance over the past year or two?
    \item Using the same 0-to-10 scale, how would you rate your overall job performance on the days you worked during the past 4 weeks (28 days)?
    \item How would you compare your overall job performance on the days you worked during the past 4 weeks (28 days) with the performance of most other workers who have a similar type of job? 
    \begin{itemize}
        \item You were a lot better than other workers
        \item You were somewhat better than other workers
        \item You were a little better than other workers
        \item You were about average
        \item You were a little worse than other workers
        \item You were somewhat worse than other workers
        \item You were a lot worse than other workers
    \end{itemize}
\end{itemize}

\subsection{The WHO-5 instrument}\label{app:who-5-instr}

The 5-item World Health Organization Well-Being Index (WHO-5, see Tab.~\ref{tab:who5-instr}) is a short and generic global rating scale measuring subjective well-being. Because the WHO considers positive well-being to be another term for mental health~\cite{jahoda}, the WHO-5 only contains positively phrased items, and its use is recommended by~\cite{bech1999health}. 

Instruction:
\begin{quote}
    Please indicate for each of the five statements which is closest to how you have been feeling over the last week from ``At no time'' (1) to ``All of the time'' (6). Over the last week: 
\end{quote}

\begin{table}[ht!]
    \centering
    \begin{tabular}{p{7cm}|c|c|c|c|c|c}
         & 1 & 2 & 3 & 4 & 5 & 6 \\\hline\hline
        I have felt cheerful and in good spirits. & & & & & & \\\hline
        I have felt calm and relaxed. & & & & & & \\\hline
        I have felt active and vigorous. & & & & & & \\\hline
        I woke up feeling fresh and rested. & & & & & & \\\hline
        My daily life has been filled with things that interest me. & & & & & & \\\hline
    \end{tabular}
    \caption{WHO-5}
    \label{tab:who5-instr}
\end{table}

\newpage
\section{Model designs}\label{app:modeldesigns}

\subsection{Gaussian Process model}\label{app:GP}

Below is the model specification for modeling the weekly or daily trends using a Gaussian Process.

\begin{align*}
\left[
\begin{array}{c}
\mathrm{Q}1_i \\
\vdots \\
\mathrm{Q}5_i \\
\end{array}
\right] 
&
\sim
\mathrm{Cumulative}
\left(
\left[
\begin{array}{c}
\phi_{\mathrm{Q}1,i}\\
\vdots\\
\phi_{\mathrm{Q}5,i}
\end{array}
, \mathbf{S}
\right]
\right) & \mathrm{[likelihood]} \\
\logit(\phi_{\mathrm{Q}\{1,\ldots,5\},i}) & =  \gamma_{\mathrm{\scriptscriptstyle{TIME}}[i]} + \alpha_{\mathrm{\scriptscriptstyle{ID}}[i]} & \text{[linear model]} \\
\left[
\begin{array}{c}
\gamma_1\\
\vdots\\
\gamma_n
\end{array}
\right]
& \sim \mathrm{MVNormal}
\left(
\left( 
\begin{array}{c}
0\\
\vdots\\
0
\end{array}
\right),
\mathbf{K}
\right) & \text{[prior Gaussian process]} \\
\mathbf{K}_{ij} & = \tau^2 \exp(- T^2_{ij}/2\rho^2) & \text{[covariance matrix } \mathcal{GP}\text{]}\\
\tau & \sim \mathrm{Weibull}(2,1) & \text{[prior std dev } \mathcal{GP}\text{]}\\
\rho & \sim \text{Inv-Gamma}(4,1) & \text{[prior length-scale } \mathcal{GP}\text{]}\\
\mathbf{S} & = \left( 
 \begin{array}{ccccc}
 \sigma_{\text{Q}1} & 0 & 0 & 0 & 0\\
 0 & \sigma_{\text{Q}2} & 0 & 0 & 0\\
 0 & 0 & \sigma_{\text{Q}3} & 0 & 0\\
 0 & 0 & 0 & \sigma_{\text{Q}4} & 0\\
 0 & 0 & 0 & 0 & \sigma_{\text{Q}5}\\
 \end{array}
 \right) \mathbf{R}
 \left( 
 \begin{array}{ccccc}
 \sigma_{\text{Q}1} & 0 & 0 & 0 & 0\\
 0 & \sigma_{\text{Q}2} & 0 & 0 & 0\\
 0 & 0 & \sigma_{\text{Q}3} & 0 & 0\\
 0 & 0 & 0 & \sigma_{\text{Q}4} & 0\\
 0 & 0 & 0 & 0 & \sigma_{\text{Q}5}\\
 \end{array}
 \right)
 & \text{[covariance matrix]}\\
\sigma_{\text{Q}1},\ldots, \sigma_{\text{Q}5} & \sim \text{Weibull}(2,1) & \text{[prior std dev among questions]}\\
\mathbf{R} & \sim \mathrm{LKJ}(2) & \text{[prior correlation matrix]}\\
\alpha_{\mathrm{\scriptscriptstyle{ID}}[i]} & \sim \mathrm{Normal}(\bar{\alpha},\sigma_{\mathrm{\scriptscriptstyle{ID}}}) & \text{[adaptive prior]}\\
\bar{\alpha} & \sim \mathrm{Normal}(0,2) & \text{[hyperprior avg ID]}\\
\sigma_{\mathrm{\scriptscriptstyle{ID}}} & \sim \mathrm{Weibull}(2,1) & \text{[hyperprior std dev of IDs]}\\
\end{align*}

For the weekly trend, on Line $1$ we assume a \textsf{Cumulative} likelihood where we model all questions' covariance using a covariance matrix $\mathbf{S}$. The linear model on the next line uses a $\logit$ link function as is default, and then models the time, $\gamma$, with a Gaussian Process ($\mathcal{GP}$), with a varying intercept $\alpha$ for subjects. 

Line $3$ places a multivariate normal distribution as prior for the $\mathcal{GP}$, while Lines $4$--$6$ declares a covariance matrix, a prior for the standard deviations, and a prior for the length-scale argument of the $\mathcal{GP}$.

On Line $7$ a covariance matrix is declared for $\mathbf{S}$. Then priors for the standard deviations among questions and the correlation matrix $\mathbf{R}$ are declared (Lines $8$--$9$). 

Finally, Lines $10$--$12$ declare an adaptive prior for the varying intercept among subjects, and hyperpriors for the average subject (Line $11$) and the standard deviation of subjects (final line).

For the daily trend the same model can be used. However, for the daily trend there was only one question asked. This means that the covariance between questions does not need to be modeled and, hence, Lines $7$--$9$ can be removed. Additionally, a suitable prior for the daily data concerning length-scale is $\text{Inv-Gamma}(1.6,0.1)$.

As is evident from the reproducibility package, prior predictive checks were conducted and the combination of priors were uniform on the outcome scale.

\newpage
\subsection{Dummy variable regression model}\label{app:dvrm}

Recall, that for the dummy variable regression models (DVRMs) each instrument (MAAS, SPANE, etc.) was modeled separately with the time ($t_0$ vs.\ $t_1$) used as an indicator (predictor). Four population-level effects (age, gender, occupation, and living conditions) and one group-level effect (subject) were used as predictors.

\begin{align*}
\left[
\begin{array}{c}
\mathrm{Q}1_i \\
\vdots \\
\mathrm{Q}n_i \\
\end{array}
\right] 
&
\sim
\mathrm{Cumulative}
\left(
\left[
\begin{array}{c}
\phi_{\mathrm{Q}1,i}\\
\vdots\\
\phi_{\mathrm{Q}n,i}
\end{array}
, \mathbf{S}
\right]
\right) & \mathrm{[likelihood]} \\
\mathbf{S} & = \left( 
\begin{array}{ccc}
 \sigma_{\text{Q}1} & 0 & 0\\
 0 & \ddots & 0 \\
 0 & 0 & \sigma_{\text{Q}n}
 \end{array}
 \right) \mathbf{R}
 \left( 
 \begin{array}{ccc}
 \sigma_{\text{Q}1} & 0 & 0\\
 0 & \ddots & 0 \\
 0 & 0 & \sigma_{\text{Q}n}
 \end{array}
 \right)
 & \text{[covariance matrix]}\\
 \sigma_{\text{Q}1},\ldots, \sigma_{\text{Q}n} & \sim \text{Weibull}(2,1) & \text{[prior std dev among questions]}\\
\mathbf{R} & \sim \mathrm{LKJ}(2) & \text{[prior correlation matrix]}\\
\logit(\phi_{\mathrm{Q}\{1, \ldots, n\},i}) & = 
\alpha \cdot \text{AGE} + 
\gamma \cdot \text{GENDER} + 
\omega \cdot \text{OCCUPATION} \\ 
& + 
\lambda \cdot \text{LIVING} +
\tau \cdot \text{TIME} +
\iota_{\mathrm{\scriptscriptstyle{ID}}[i]} & \text{[linear model]}\\
\alpha, \gamma, \omega, \lambda, \tau & \sim \mathrm{Normal}(0,3) & \text{[priors population-level effects]}\\
\iota_{\mathrm{\scriptscriptstyle{ID}}[i]} & \sim \mathrm{Normal}(\bar{\alpha},\sigma_{\mathrm{\scriptscriptstyle{ID}}}) & \text{[adaptive prior]}\\
\bar{\alpha} & \sim \mathrm{Normal}(0,2) & \text{[hyperprior avg ID]}\\
\sigma_{\mathrm{\scriptscriptstyle{ID}}} & \sim \mathrm{Weibull}(2,1) & \text{[hyperprior std dev of IDs]}
\end{align*}

For each instrument we assumed a \textsf{Cumulative} likelihood where all questions' covariance was modeled by a covariance matrix $\mathbf{S}$. On Line $2$ the covariance matrix is declared for $\mathbf{S}$ and priors for the standard deviations among questions and the correlation matrix $\mathbf{R}$ are declared on Lines $3$--$4$). 

The linear model on the next two lines uses a $\logit$ link function as is default, and then declares five population-level parameters and a varying intercept $\iota$ for subjects. On Line $7$ priors for the population-level parameters are declared.

Finally, Lines $8$--$10$ an adaptive prior with hyperpriors is declared for the varying intercept $\iota$.

The only thing that differs between the instruments are the number of questions asked. This implies that the covariance matrix $\mathbf{S}$ differs in size depending on number of questions.

Additionally, for one instrument, SE, there were two questions modeled with a $\mathsf{Bernoulli}$ likelihood due to responses on two levels.

As is evident from the reproducibility package, prior predictive checks were conducted and the combination of priors were uniform on the outcome scale.

\newpage
\section{Detailed Findings: Significant Effects of Other Predictors}\label{app:detailedfindings}

\begin{figure}[h]
    \centering
    \includegraphics[width=0.8\textwidth]{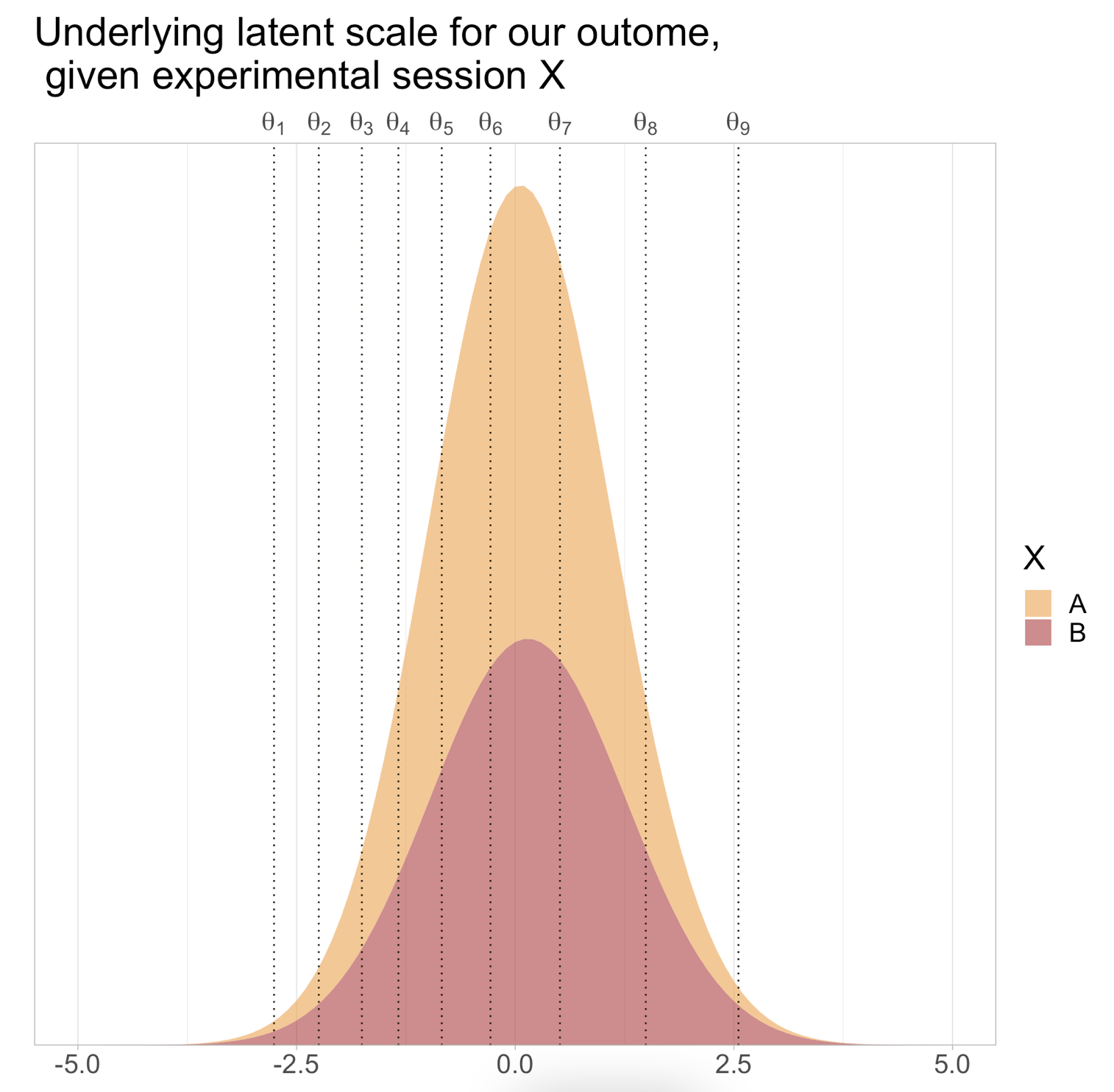}
    \caption{Underlying latent scale for outcome, given experimental session X}
    \label{fig:latentscale}
\end{figure}

To show that the experiments of run 1 and run 2 confirm the general tendencies, we confirm the underlying latent scale in Fig.~\ref{fig:latentscale}. The similar curves with similar centers of the peak show that there is no threat to validity given by the two different lengths of the experiment. In addition, combining the two runs gives the model more certainty, which makes the results more reliable. Had we taken the results of both runs separately, there would be more uncertainty in both individual models, but this was not necessary given the present latency.

\subsection{Mindfulness Attention Awareness Scale}
The MAAS instrument (App.~\ref{app:maas-instr}) consisted of $15$ statements to agree or disagree with. Eleven of the ratings indicated a significant difference at $t_0$ vs.\ $t_1$: Q$1$--$8$, $11$--$12$, and $14$. 
\noindent
In all the above cases the effect was negative, i.e., the responses were higher at $t_0$ than at $t_1$ (please see Fig.~\ref{fig:mcmc_MAAS}. If we look at the other predictors, \textsf{age} and \textsf{gender} did not have a significant effect, while \textsf{occupation} was significant (negative) for Q$2$, i.e., ``I break or spill things because of carelessness, not paying attention, or thinking of something else.''

\begin{figure*}
    \centering
    \includegraphics[width=\textwidth]{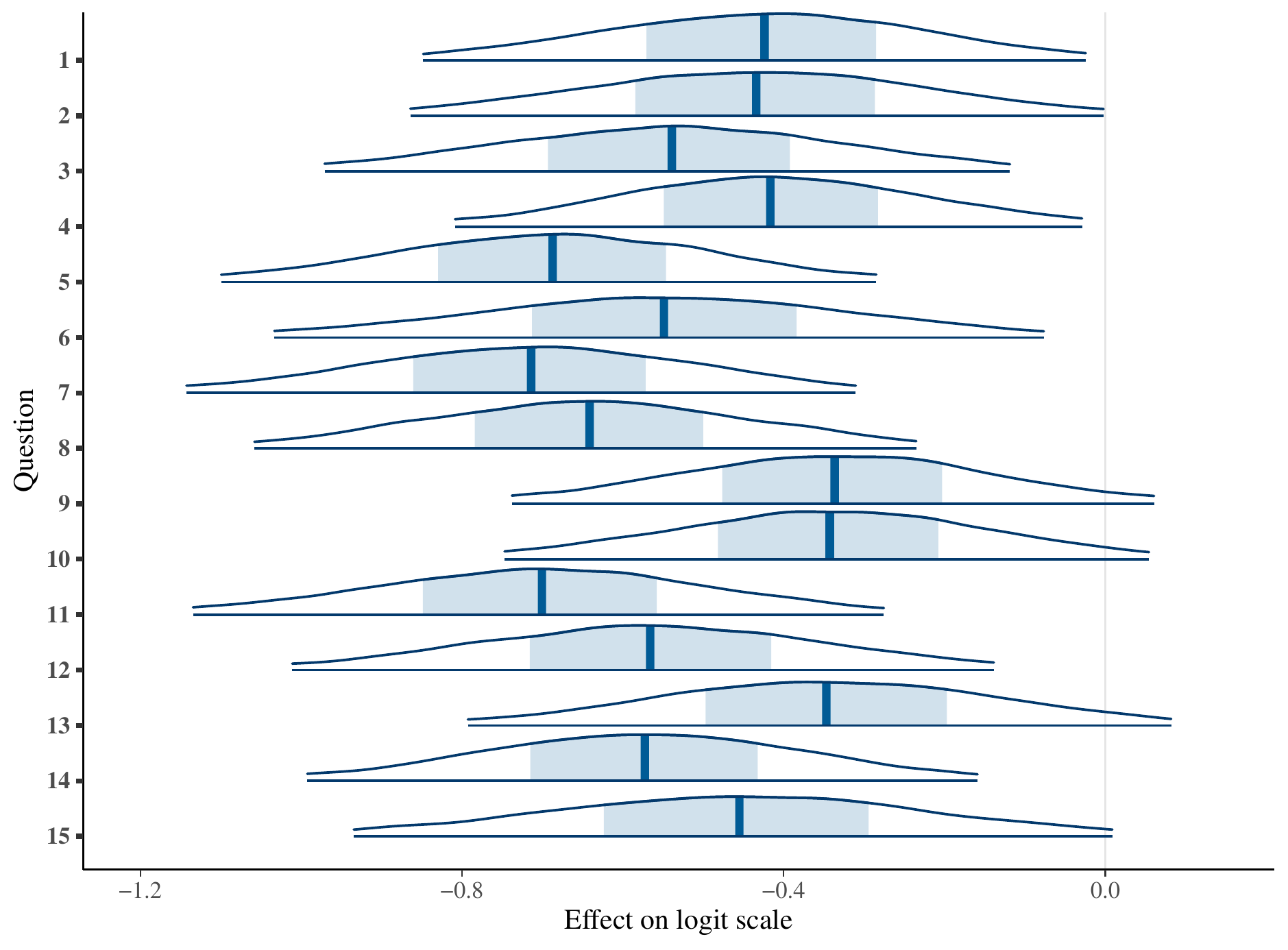}
    \caption{MAAS Density plots computed from posterior draws. The densities are cut off at 95\% and the shaded area is the 50\% uncertainty interval. We can see a number of questions not crossing zero (no effect observed).}
\end{figure*}

Additionally, the predictor \textsf{living condition} was significant (negative) in Q$1$--$3$, $8$, and $12$ (items listed in App.~\ref{app:maas-instr}).

\subsection{Scale of Positive And Negative Experiences}

For the SPANE items, see App.~\ref{app:spane-pwb-pts-instr}.
The results for the predictor time are in Fig.~\ref{fig:spane-effects}.

Below we summarize the significant effects of the other predictors.
In all the following tables for predictors, a $+$ means that the item was rated higher for that variable, and a $-$ means that the item was rated lower for that variable.\\
For gender, a $-$ means that females rated themselves more negatively than males, and a $+$ means that females rated themselves more positively. This is not visible directly from the table below, but requires to know how the data was coded inside the model. For this specific reason, we moved these tables into the appendix, as they are not relevant to understand the narrative of the article, but can be considered interesting observations.

\begin{tabularx}{\textwidth}{r l l l X}
    \hline
    Question & Age & Gender & Occupation & Living conditions \\
    \hline
    Q$3$ & & $-$ & & \\
    Q$6$ & & $-$ & & \\
    Q$7$ & & $-$ & & \\
    Q$9$ & $+$ & & &\\
    \hline
\end{tabularx}
\noindent In summary for this table, the higher the \textsf{age}, the higher the response in Q$9$. Concerning \textsf{gender}, males answered with higher values in Q$3$, Q$6$, and Q$7$.

\subsection{Psychological Well-Being}

\begin{figure}
    \centering
    \includegraphics[width=0.8\textwidth]{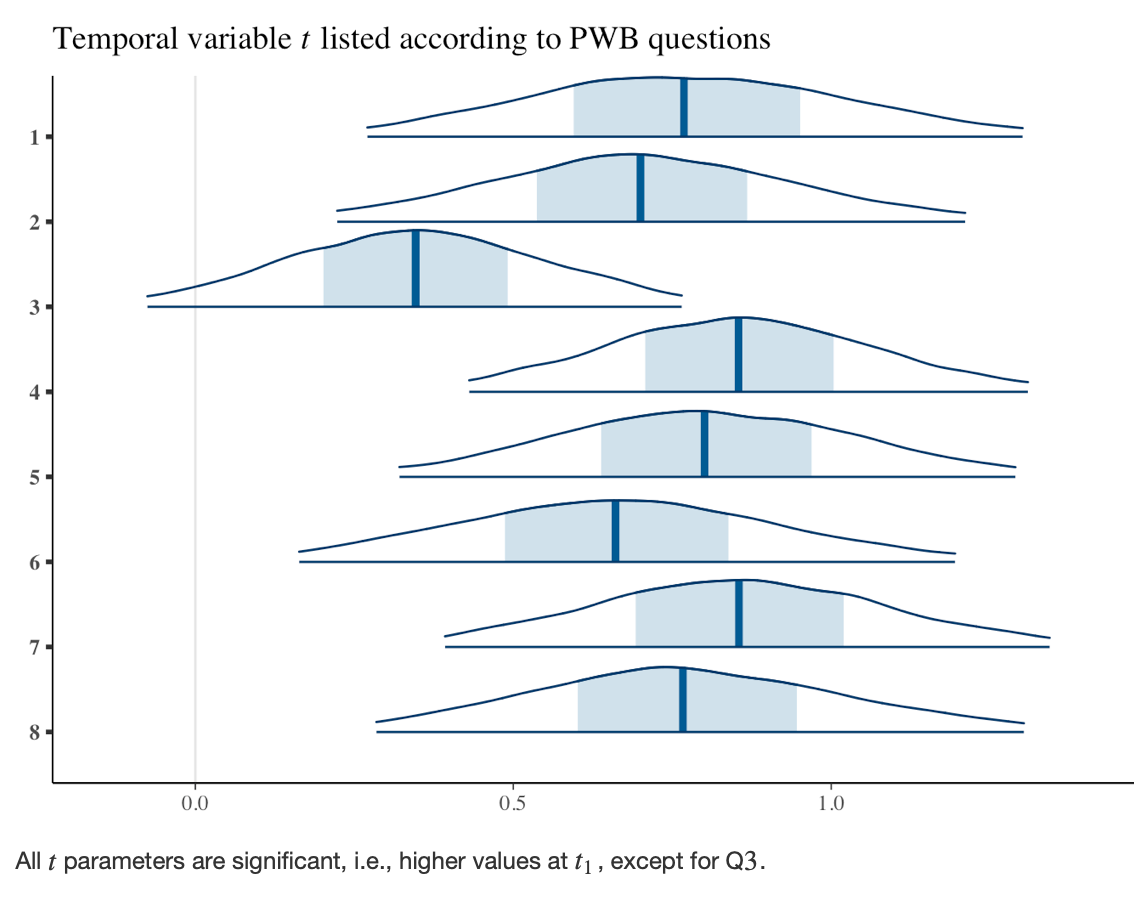}
    \caption{The effects of $t$ for the PWB instrument. 
    The temporal variable $t$ clearly has an effect (positive) in all questions except Q3.}
    \label{fig:pwb-effects}
\end{figure}

Figure~\ref{fig:pwb-effects} shows the effects for the predictor time. The temporal variable $t$ clearly has an effect (positive) in all questions except Q3.

Below we summarize the significant effects of the other predictors for PWB (for the items, see App.~\ref{app:spane-pwb-pts-instr}). The same logic applies here as in the previous table; however, one new effect is present, i.e, \textsf{occupation}. In Q$3$ (\textit{I am engaged and interested in my daily activities.}), participants with occupation \textsf{student} replied with \textit{higher} responses compared to others.

\begin{tabularx}{\textwidth}{r l l l X}
    \hline
    Question & Age & Gender & Occupation & Living conditions \\
    \hline
    Q$1$ & & & & $+$ \\
    Q$2$ & $-$ & $-$ & & \\
    Q$3$ & $+$ & & $-$ & \\
    Q$4$ & & $-$ & & \\
    Q$7$ & & & & $-$\\
    \hline
\end{tabularx}

\subsection{Positive Thinking Scale}

For the PTS items, see App.~\ref{app:spane-pwb-pts-instr}.
The results for the predictor time are given below in Fig.~\ref{fig:pts-effects}.

\begin{figure}
    \centering
    \includegraphics[width=\textwidth]{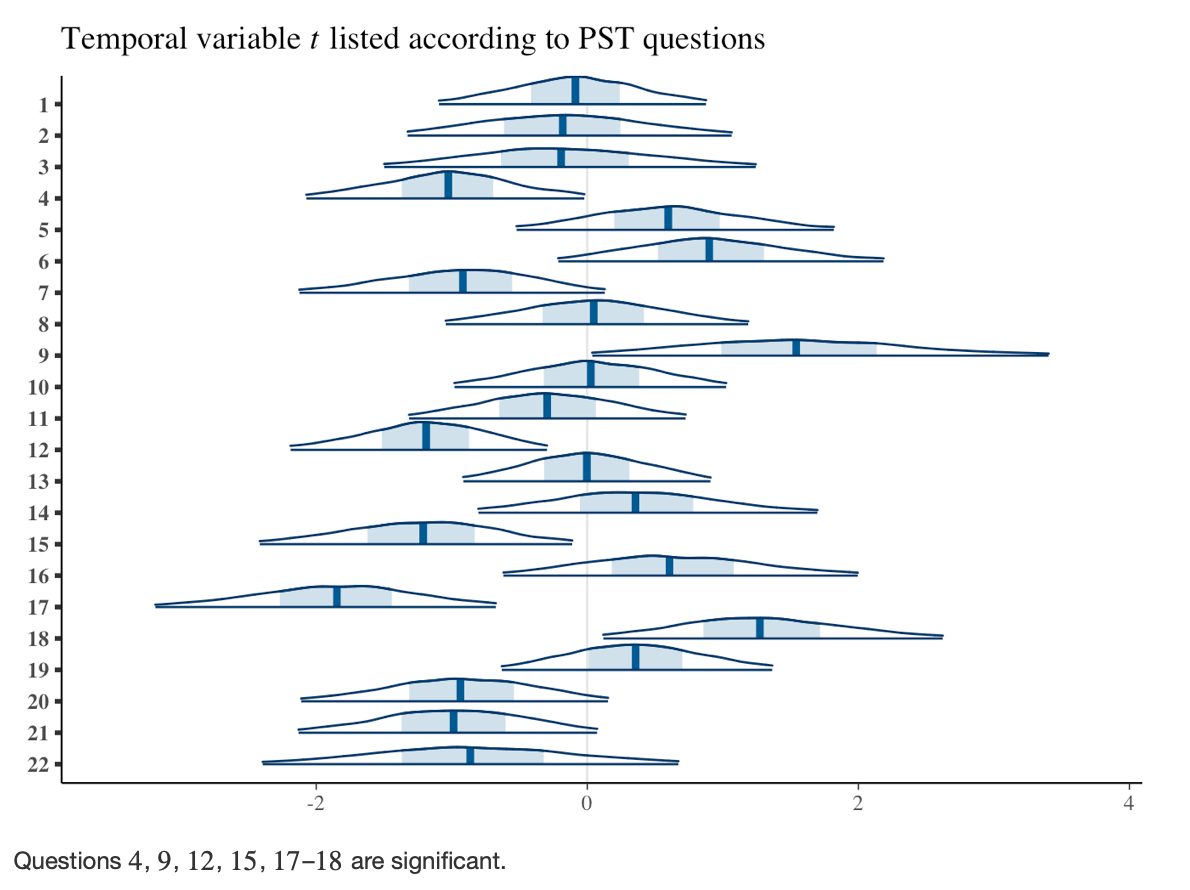}
    \caption{The PTS results for the predictor time.}
    \label{fig:pts-effects}
\end{figure}

Below we summarize the significant effects of the other predictors. Please refer to the appendix for the respective survey items.

\begin{tabularx}{\textwidth}{r l l l X}
    \hline
    Question & Age & Gender & Occupation & Living conditions \\
    \hline
    Q$1$ & & & $-$ & $-$\\
    Q$3$ & & & $+$ & \\
    Q$11$ & $-$ & $-$ & & \\
    Q$16$ & & & & $+$\\
    Q$17$ & & & $-$ & $-$\\
    Q$19$ & & & & $-$\\
    \hline
\end{tabularx}

\subsection{Self Efficacy}
\begin{figure}
    \centering
    \includegraphics[width=0.8\textwidth]{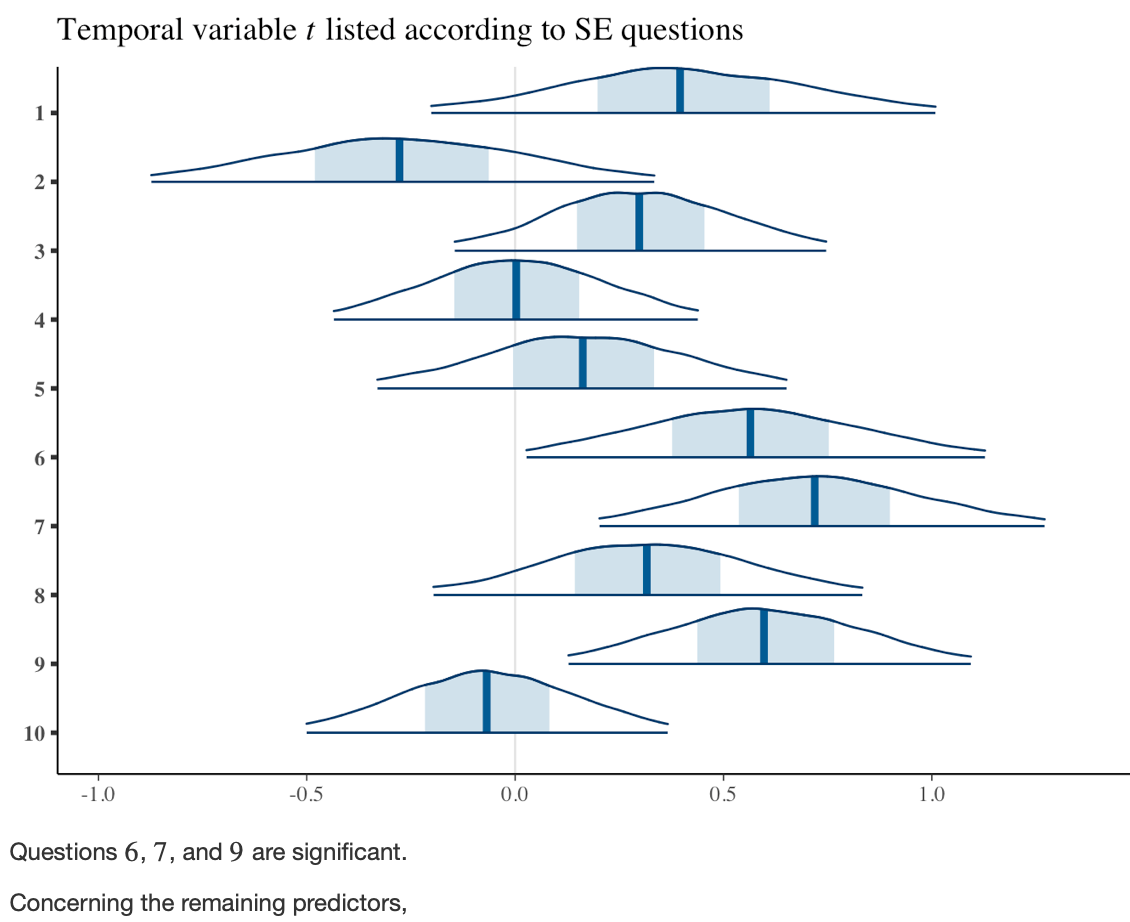}
    \caption{SE effects for predictor time.}
    \label{fig:se-effects}
\end{figure}

The SE instrument (App.~\ref{app:se-instr}) consisted of ten questions (Likert $1$--$4$). Questions $6$, $7$, and $9$ showed a significant effect (positive), i.e., higher responses at $t_1$, see Fig.~\ref{fig:se-effects}.

\begin{itemize}
    \item[Q$6$] I can easily face difficulties because I can always trust my abilities.
    \item[Q$7$] Whatever happens, I'll be fine.
    \item[Q$9$] When a new thing comes to me, I know how to handle it.
\end{itemize}

Concerning the other predictors, no significant effects were present, i.e., it is not clear which predictors drove the significant difference between $t_0$ and $t_1$.

\subsection{Perceived Productivity}

\begin{figure}
    \centering
    \includegraphics[width=0.8\textwidth]{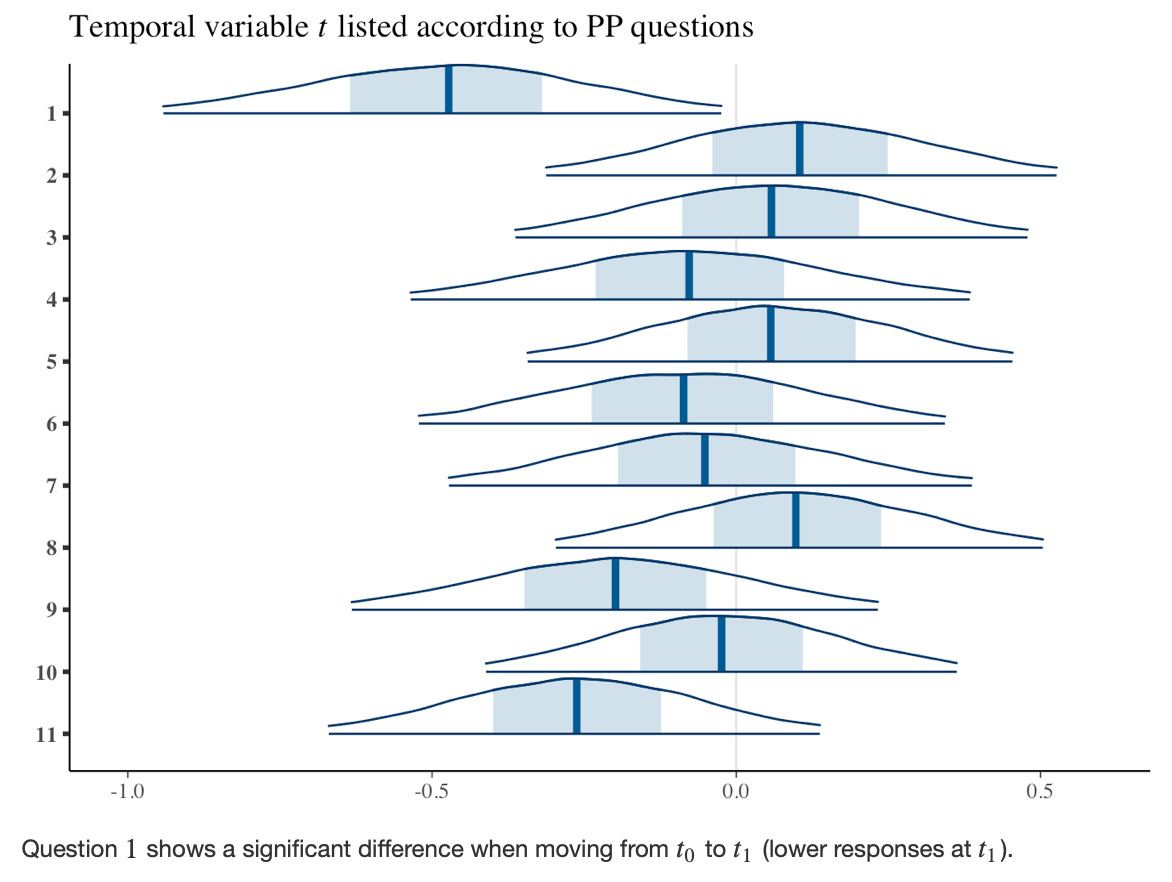}
    \caption{The PP results for the predictor time.}
    \label{fig:pp-effects}
\end{figure}

The HPQ part consisted of eleven questions (with Likert scales varying, going up to $5$, $7$, or $10$, depending on the question, see App.~\ref{app:pp-instr}).  The results for the predictor time are given in Fig.~\ref{fig:pp-effects}. Only Q$1$ (\textit{How often was your performance higher than most workers on your job?}) shows a significant difference when moving from $t_0$ to $t_1$ (lower responses at $t_1$).

Below we summarize the significant effects of the other predictors, i.e. Q$3$ (\textit{How often did you do no work at times when you were supposed to be working?}) showing a higher score for gender \textsf{female}, and Q$5$ (\textit{How often was the quality of your work lower than it should have been?}) showing a lower score when the living condition was \textsf{shared} with partner or family as opposed to living by oneself.

\begin{tabularx}{\textwidth}{r l l l X}
    \hline
    Question & Age & Gender & Occupation & Living conditions \\
    \hline
    Q$3$ & & $+$ & & \\
    Q$5$ & & & & $-$\\
    \hline
\end{tabularx}

\subsection{Predictor Number of Sessions}\label{sec:app:det-sess-numbers}

The following Table~\ref{tab:totalnumbersess} shows an overview of all significant effects for total number of sessions as predictor.
The first column is an ID, the rowname indicates the variable of the instrument, e.g. MAASQ116\_total\_sessions refers to MAAS question 1 (Likert scale 1 -6) for total sessions attended.
The next two columns indicate the estimate and the estimation error.
Please note that for SPANE, the results seem to be alternating, but looking back at the instrument (see Sec.~\ref{app:spane-pwb-pts-instr}), half of the items were scored reversely in exactly the pattern that is reflected here.

\begin{table}[h]
    \centering
    \begin{tabular}{c|c|c|c|c|c}
                          &  rowname &  Estimate & Est.Error  &     Q2.5   &    Q97.5 \\\hline
1  & $MAASQ116_total_sessions$ & -0.3155496 & 0.1216486 & -0.5569970 & -0.08184895 \\
2 &  $MAASQ216_total_sessions$ & -0.3804576 & 0.1273390 & -0.6411205 & -0.13837388 \\
3 &  $MAASQ316_total_sessions$ & -0.2634123 & 0.1231561 & -0.5100075 & -0.02042572 \\
5 &  $MAASQ516_total_sessions$ & -0.3689709 & 0.1167109 & -0.6023460 & -0.14523413 \\
6 &  $MAASQ616_total_sessions$ & -0.2894895 & 0.1305140 & -0.5477286 & -0.03658702 \\
7 &  $MAASQ716_total_sessions$ & -0.3647491 & 0.1231923 & -0.6058283 & -0.12399760 \\
8 &  $MAASQ816_total_sessions$ & -0.2611191 & 0.1214209 & -0.5011597 & -0.02438610 \\
10 & $MAASQ1016_total_sessions$ & -0.2886498 & 0.1174016 & -0.5226733 &  -0.05928175 \\
11 & $MAASQ1116_total_sessions$ & -0.4540885 &  0.1211362 & -0.6957715 & -0.21564968 \\
12 & $MAASQ1216_total_sessions$ & -0.2509503 &  0.1246984 & -0.4957514 & -0.01287479 \\
14 & $MAASQ1416_total_sessions$ & -0.4358311 &  0.1179180 & -0.6693166 & -0.20832100 \\\hline
1  & $SPANEQ115_total_sessions$  & 0.4662730 & 0.1511756 & 0.1756023 & 0.77767102 \\
2 &  $SPANEQ215_total_sessions$ & -0.5187067 & 0.1341723 & -0.7911272 & -0.25860345 \\
3 &  $SPANEQ315_total_sessions$  & 0.4918396 & 0.1524530 & 0.2054288 & 0.80508272 \\
4 &  $SPANEQ415_total_sessions$ & -0.4509748 & 0.1308125 & -0.7134680 & -0.20059290 \\
5 &  $SPANEQ515_total_sessions$ & 0.3955807 & 0.1311677 & 0.1416188 & 0.65872865\\
6 &  $SPANEQ615_total_sessions$ & -0.2643148 & 0.1243299 & -0.5096883 & -0.01981721 \\
7 &  $SPANEQ715_total_sessions$ & 0.5689896 & 0.1411704 & 0.3003263 & 0.84980113 \\
8 &  $SPANEQ815_total_sessions$ & -0.3191885 & 0.1221512 & -0.5628583 & -0.08297879 \\
9 &  $SPANEQ915_total_sessions$ & -0.4594716 & 0.1445001 & -0.7530686 & -0.18126877 \\
10 & $SPANEQ1015_total_sessions$ & 0.3753050 & 0.1239305 & 0.1374020 & 0.61918997 \\
11 & $SPANEQ1115_total_sessions$ & -0.2855759 & 0.1255116 & -0.5319962 & -0.04022932 \\\hline
1 & $PWBQ117_total_sessions$ & 0.3232594 & 0.1505071 & 0.03253444 & 0.6231016 \\
2 & $PWBQ217_total_sessions$ & 0.2971393 & 0.1408516 & 0.02316987 & 0.5816784 \\
4 & $PWBQ417_total_sessions$ & 0.3391010 & 0.1257622 & 0.09843479 & 0.5881914 \\
5 & $PWBQ517_total_sessions$ & 0.2659871 & 0.1345689 & 0.01074883 & 0.5332659 \\
6 & $PWBQ617_total_sessions$ & 0.3150326 & 0.1478417 & 0.02922867 & 0.6087226 \\
7 & $PWBQ717_total_sessions$ & 0.3061679 & 0.1298780 & 0.05548536 & 0.5639220 \\
8 & $PWBQ817_total_sessions$ & 0.3378056 & 0.1378315 & 0.07209965 & 0.6095512 \\\hline
9 & $PSTQ901_total_sessions$ & 1.9809234 & 0.9627086  & 0.470679200  & 4.20554675 \\
12 & $PSTQ1201_total_sessions$ & -0.5350738 & 0.2776467 & -1.103002250 & -0.01894271 \\
17 & $PSTQ1701_total_sessions$ & -0.9643101 & 0.3736942 & -1.751491250 & -0.28321947 \\
18 & $PSTQ1801_total_sessions$ & 0.6554668 & 0.3499538 & 0.009657123 & 1.38352275 \\\hline
7 & $SEQ714_total_sessions$ & 0.4327188 & 0.1503527 & 0.1492953 & 0.736231 \\\hline
6 & $PPHQ615_total_sessions$ & -0.2617817 & 0.1271908 & -0.5070381 & -0.01792041 \\\hline
    \end{tabular}
    \caption{Significant effects for total number of sessions as predictor}
    \label{tab:totalnumbersess}
\end{table}

The following Table~\ref{tab:numbersessliverecorded} shows an overview of all significant effects for number of sessions live and recorded as predictor.

\begin{table}[h]
    \centering
    \begin{tabular}{c|c|c|c|c|c}
                             &  rowname &  Estimate Est. & Error   &    Q2.5    &   Q97.5 \\\hline
1 &      $MAASQ116_live_sessions$ & -0.2939231 & 0.1254551 & -0.5422444 &  -0.04608940 \\
3 &       $MAASQ216_live_sessions$ & -0.2663390 & 0.1292418 & -0.5273328 & -0.01924983 \\
9 &      $MAASQ516_live_sessions$ & -0.3714334 & 0.1238394 & -0.6162243 & -0.13673993 \\
13 &     $MAASQ716_live_sessions$ & -0.3463737 & 0.1278613 & -0.5996803 & -0.10126535 \\
19 &    $MAASQ1016_live_sessions$ & -0.2683671 &  0.1265510 & -0.5189259 & -0.02028169 \\
21  &   $MAASQ1116_live_sessions$ & -0.3519825 & 0.1277744 & -0.6044776 & -0.10675305 \\
22 & $MAASQ1116_recorded_sessions$ & -0.2729695 & 0.1309911 & -0.5282038 & -0.01542422 \\
27 &   $MAASQ1416_live_sessions$ & -0.4692584 & 0.1242340 & -0.7120738 & -0.22517830 \\\hline
1 &  $SPANEQ115_live_sessions$  & 0.3714696 & 0.1564535 & 0.07600304 & 0.68959905 \\
3 &  $SPANEQ215_live_sessions$ & -0.4590504 & 0.1376351 & -0.73415957 & -0.19549265 \\
5 &   $SPANEQ315_live_sessions$ & 0.3926880 & 0.1523264 & 0.10560065 & 0.70054970 \\
7 &  $SPANEQ415_live_sessions$ & -0.3858643 & 0.1326585 & -0.65485220 & -0.13025333 \\
9  & $SPANEQ515_live_sessions$ & 0.4090131 & 0.1388661 & 0.14721700 & 0.69743993 \\
13 & $SPANEQ715_live_sessions$ & 0.6621751 & 0.1569447 & 0.36908735 & 0.98411335 \\
15 & $SPANEQ815_live_sessions$ & -0.2616480 & 0.1268337 & -0.51337150 & -0.01369651 \\
17 & $SPANEQ915_live_sessions$ & -0.3424188 & 0.1480671 & -0.63512853 & -0.05596512 \\
19 & $SPANEQ1015_live_sessions$ & 0.4365976 & 0.1324461 & 0.17729748 & 0.69863018 \\\hline
7  &    $PWBQ417_live_sessions$ & 0.3380077 & 0.1354551 & 0.07777354 & 0.6092814 \\
9 &     $PWBQ517_live_sessions$ & 0.2920369 & 0.1449304 & 0.01599527 & 0.5898144 \\
16 & $PWBQ817_recorded_sessions$ & 0.3250622 & 0.1521206 & 0.03104731 & 0.6273903 \\\hline
2 &  $PSTQ101_recorded_sessions$ & -0.8114198 & 0.4127656 & -1.70676300 & -0.08103662 \\
8 &  $PSTQ401_recorded_sessions$ & -0.6589501 & 0.3618867 & -1.41424650 & -0.01133453 \\
17 &     $PSTQ901_live_sessions$ & 3.1336475 & 1.5388280 & 0.76178103 & 6.67567475 \\
22 & $PSTQ1101_recorded_sessions$ & -1.2048545 & 0.4597776 & -2.20838500 & -0.40500393 \\
28 & $PSTQ1401_recorded_sessions$ & 1.5137920 & 0.9497095 & 0.02391954 & 3.70106175 \\
33 &    $PSTQ1701_live_sessions$ & -0.7760812 & 0.3848592 & -1.59662025 & -0.07400772 \\
36 & $PSTQ1801_recorded_sessions$ & 1.0777616 & 0.6296321 & 0.02372501 & 2.47252725 \\
44 & $PSTQ2201_recorded_sessions$ & 2.0895201 & 1.2682835 & 0.10201318 & 4.99020650 \\\hline
13 & $SEQ714_live_sessions$ & 0.4491137 & 0.158366 & 0.1545877 & 0.7680788 \\\hline
22 & $PPOQ117_recorded_sessions$ & -0.2526645 & 0.1271695 & -0.5066494 & -0.003687787 \\\hline
    \end{tabular}
    \caption{Significant effects for number of live and recorded sessions as predictor}
    \label{tab:numbersessliverecorded}
\end{table}

\end{document}